\documentclass[12pt]{article}
\usepackage{amsmath}
\usepackage{prodint}
\usepackage{graphicx}
\usepackage{enumerate}
\usepackage{natbib}
\usepackage{url} 
\usepackage{binomexp}
\usepackage{mathtools}
\usepackage{amsfonts}
\usepackage{mathrsfs}
\RequirePackage{subcaption}
\usepackage{booktabs}
\usepackage{bbm}
\usepackage{dsfont}
\usepackage{longtable}
\usepackage{yhmath}
\usepackage{bm}
\usepackage{amssymb}
\usepackage[nodisplayskipstretch]{setspace}
\usepackage{tikz}
\usepackage{float}
\usepackage{color}

\usepackage{tabularx}
\usepackage{array}
\newcolumntype{L}{>{\raggedright\arraybackslash}X}

\usepackage{stmaryrd}
\usepackage{paralist}
\usepackage{multirow}
\usepackage{threeparttable}
\usepackage{soul}
\usepackage{algorithm,algpseudocode}
\makeatletter
\newcommand{\algmargin}{\the\ALG@thistlm}
\makeatother
\newlength{\whilewidth}
\algdef{SE}[parWHILE]{parWhile}{EndparWhile}[1]
  {\parbox[t]{\dimexpr\linewidth-\algmargin}{%
     \hangindent\whilewidth\strut\algorithmicwhile\ #1\ \algorithmicdo\strut}}{\algorithmicend\ \algorithmicwhile}%
\algnewcommand{\parState}[1]{\State%
  \parbox[t]{\dimexpr\linewidth-\algmargin}{\strut #1\strut}}

\newcommand{\bfx}{\bm{x}}

\newcommand{\bftheta}{\bm{\theta}}

\newcommand{\RR}{{\mathrm{RR}}}
\newcommand{\BF}{{\mathrm{BF}}}

\newcommand{\bbE}{\mathbb{E}}

\newcommand{\bbP}{\mathbb{P}}

\newcommand{\bbone}{\mathbbm{1}}

\newcommand{\calS}{\mathcal{S}}

\newcommand{\calX}{\mathcal{X}}
\newcommand{\calY}{\mathcal{Y}}

\usepackage[colorlinks=true,linkcolor=red,citecolor=black]{hyperref}

\newtheorem{theorem}{Theorem}
\newtheorem{assumption}{Assumption}
\newtheorem{proposition}{Proposition}
\newtheorem{corollary}{Corollary}
\newtheorem{lemma}{Lemma}

\newtheorem{example}{Example}

\newcommand{\blind}{1}

\allowdisplaybreaks

\usetikzlibrary{shapes,decorations,arrows,calc,arrows.meta,fit,positioning}
\tikzset{
     -Latex,auto,node distance =1 cm and 1 cm,semithick,
     state/.style ={ellipse, draw, minimum width = 0.7 cm},
     point/.style = {circle, draw, inner sep=0.04cm,fill,node contents={}},
    bidirected/.style={Latex-Latex,dashed},
     el/.style = {inner sep=2pt, align=left, sloped}
 }

\begin{document}

\def\spacingset#1{\renewcommand{\baselinestretch}%
{#1}\small\normalsize} \spacingset{1}


\if1\blind
{
    \title{\Large \bf Beyond principal ignorability: Nonparametric sensitivity bounds for principal stratification}
    \author{Xinyuan Chen$^{1,\ast}$ ~~ Michael O. Harhay$^2$ ~~ Fan Li$^3$\vspace{0.2cm}\\
    $^1$Department of Mathematics and Statistics,\\ Mississippi State University, MS, USA\\
    $^2$Department of Biostatistics, Epidemiology, and Informatics, \\
    University of Pennsylvania, Philadelphia, PA, USA\\
    $^3$Department of Biostatistics, Yale School of Public Health, CT, USA\\
    ${}^\ast$xchen@math.msstate.edu}  
  \maketitle
} \fi

\if0\blind
{
  \bigskip
  \bigskip
  \bigskip
  \begin{center}
    {\Large \bf Beyond principal ignorability: Nonparametric sensitivity bounds for principal stratification}
\end{center}
  \medskip
} \fi

\bigskip
\begin{abstract}
	Principal stratification is an effective framework addressing intermediate variables in causal inference. However, point identification of the principal causal effects (PCEs) often requires the untestable principal ignorability (PI) assumption. This article develops a nonparametric sensitivity analysis framework for evaluating PI violations. We introduce a margin-free bounding factor parameterized by the selection and outcome relative risks of an unmeasured confounder. Using this bounding factor, we derive sharp nonparametric bounds for each PCE. We prove that these bounds nest within the worst-case nonparametric bounds with and without the monotonicity assumption. We then discuss Cornfield-type conditions and principal E-values that quantify the minimum joint magnitude of unmeasured confounding required to nullify the target PCE. Furthermore, we generalize this methodology to principal generalized causal effects, extending the sensitivity bounds and falsification thresholds to the recent pairwise comparison estimands evaluated over a product space.
\end{abstract}

\noindent%

\spacingset{1.75} 

\section{Introduction}

Principal stratification \citep{Frangakis2002} is an effective framework to address intermediate variables (e.g., non-compliance, truncation by death) when estimating causal effects. This framework partitions the population into principal strata, each characterized by the ordered pair of potential values that the intermediate variables would assume under different treatment conditions. The partition leads to a mixture of latent strata in each observed data cell defined by the treatment and the intermediate variable values, resulting in partial identification of the target estimands, the principal causal effects (PCEs). To attain point identification, an increasingly adopted structural assumption is principal ignorability (PI) \citep{Ding2016}, which posits that the potential final outcome and the latent principal stratum membership are conditionally independent given the observed covariates. As this assumption is fundamentally untestable, sensitivity analyses that yield partial identification bounds on the PCEs under departures from PI are of strong interest. 

Existing sensitivity analyses under principal stratification primarily follow three strategies: (i) nonparametric large-sample bounds that may be further tightened via relatively weak distributional assumptions \citep[e.g.,][]{Zhang2003, Imai2008, Grilli2008}; (ii) the specification of Bayesian priors on latent, unobserved quantities \citep[e.g.,][]{Imbens1997, Schwartz2011}; and (iii) the imposition of additional structural constraints. The last strategy usually involves a monotonicity assumption, which reduces the number of principal strata \citep[e.g.,][]{Long2013}, frequently in combination with exclusion restriction that constrains the value of PCEs of certain strata \citep[e.g.,][]{Miratrix2018, Lu2018} or with secondary outcomes \citep[e.g.,][]{Yin2018}. Some approaches instead exploit the exclusion restriction primarily to derive bounds on the average treatment effect (ATE), rather than on the PCEs \citep[e.g.,][]{Levis2025}.

In this article, we revisit sensitivity analysis for unverifiable assumptions under principal stratification, but approach this question through the lens of the PI assumption. By investigating departures from the PI assumption, we aim to generate actionable, sharp, nonparametric bounds on the PCEs without relying on parametric assumptions, Bayesian priors, monotonicity, or exclusion restriction. Our approach draws upon the seminal confounder-based framework of \citet{Ding2016epi} and \citet{Ding2016bmk} by deriving the suitable bounding factor, a margin-free sensitivity parameter that quantitatively characterizes the degree of PI violation. However, our nonparametric sensitivity analysis under principal stratification differs from \citet{Ding2016epi} and \citet{Ding2016bmk} in two key respects. First, the target latent stratum mean is not directly observed. The observed mean by each treatment level and intermediate outcome value is a weighted mixture of contributions from the target principal stratum and a companion, nuisance principal stratum, with the mixing weights determined by the principal scores. This mixture structure means that the existing bounding factor cannot be applied directly to the observable cell means but must instead propagate through the mixture decomposition. Second, the cross-world dependence between the two potential intermediate outcomes introduces an additional sensitivity dimension that governs the latent stratum composition itself. Taken together, these two features demand a new derivation rather than a routine application of existing results.

Leveraging our nonparametric sensitivity analysis results, we further prove that, for any finite bounding factor, the proposed bounds nest within those of \citet{Grilli2008}; and as the bounding factor tends to infinity, the proposed bounds converge to those of \citet{Grilli2008}, which are the ``worst-case'' nonparametric bounds; hence, our results serve as a bridge to unify existing bounds. Under the monotonicity assumption, we establish an analogous relationship between the proposed bounds and those of \citet{Long2013}. Importantly, leveraging Cornfield-type conditions \citep{Cornfield2009, Ding2014}, the proposed method yields two critical falsification thresholds that nullify PCEs across both additive and multiplicative scales. 

Next, to further expand on the utility of the proposed framework, we offer a generalization of our theoretical results to address the principal general causal effects (PGCEs) \citep{Chen2026}, a more general class of causal effects defined with general nonlinear contrast functions that apply to a wider variety of final outcomes, e.g., the principal probabilistic index estimand. In addition to bounds from the proposed framework, we give worst-case nonparametric bounds of the types introduced by \citet{Grilli2008} and \citet{Long2013} on the PGCEs and clarify their relationships. A novel set of Cornfield-type conditions, with corresponding falsification thresholds that nullify PGCEs, is also provided. For ease of reference, Table \ref{tab:comparison} presents a comparison between the proposed approach and several existing sensitivity methods for principal stratification.

\begin{table}[htbp]
    \centering
    \caption{A conceptual comparison with existing sensitivity analysis methods for principal stratification.}
    \label{tab:comparison}
    \footnotesize
    \setlength{\tabcolsep}{4pt}
    \renewcommand{\arraystretch}{1.3}
    \begin{tabularx}{\textwidth}{l L L L L L}
        \toprule
        \textbf{Method} & \textbf{Identifying strategy} & \textbf{Sensitivity dim.} & \textbf{Assum. \& Prior} & \textbf{Target} & \textbf{Falsif.} \\
        \midrule
        \citet{Imbens1997} & Bayesian priors & Prior parameters & Bayesian priors & PCE & Bayesian posterior \\
        \citet{Zhang2003} & Worst-case & None & Monot. or Ranked ave. score or None & PCE (SACE) & None \\
        \citet{Grilli2008} & Worst-case & None & None or Rel. majority $+$ Stoch. dom. & PCE & None \\
        \citet{Imai2008} & Worst-case & None & None or Monot. or Stoch. dom. & PCE (SACE \& SQCE) & None \\
        \citet{Schwartz2011} & Bayesian semipar. & Prior parameters & Bayesian priors & PCE & Bayesian posterior \\
        \citet{Long2013} & Worst-case $+$ covariates & None & Monot. & PCE & None \\
        \citet{Yin2018} & Worst-case & None & Monot. and/or Tran. $+$ Excl. restr. & PCE & None \\
        \citet{Miratrix2018} & Worst-case & None & Monot. $+$ Irre. alt. $+$ Excl. restr. & PCE & None \\
        \citet{Levis2025} & Instr. var. $+$ covariates & None & Excl. restr. & ATE & None \\
        \midrule
        \textbf{Proposed} & \textbf{PI relax. via $U$} & \textbf{BF + $\theta$ (2-D)} & \textbf{None} & \textbf{PCE + PGCE} & \textbf{Principal E-value} \\
        \bottomrule
        \multicolumn{6}{l}{$^*$SACE: survivor average causal effect; SQCE: survivor quantile causal effect.}\\
        \multicolumn{6}{l}{$^{**}$$U$, $\BF$, and $\theta$ are to be defined in detail later.}
    \end{tabularx}
\end{table}

The remainder of the article is organized as follows. Section \ref{sec:setting} introduces the setup, including notation and structural assumptions. Section \ref{sec:PCE-bounds} defines the sensitivity parameters and derives bounds on the PCEs. Section \ref{sec:PGCE-bounds} extends the theoretical results to the PGCEs. Section \ref{sec:illustration} presents the illustrations via two data examples. Section \ref{sec:discussion} concludes with discussions. Proofs of theoretical results are available in the Supplementary Materials.

\section{Notation and setup for principal stratification} \label{sec:setting}

We operate within the potential outcomes framework, and let $D(z)$ and $Y(z)$ denote the potential intermediate and final outcomes under treatment $Z=z\in\{0,1\}$, where $D(z)\in\{0,1\}$ and $Y(z)\in\calY$. To focus ideas, we restrict attention to the case where $\calY = [0,1]$. However, the proposed framework is sufficiently general to accommodate outcomes that are either lower-bounded or bounded on both sides, since one may apply a proper location-scale transformation and relabeling to ensure that $Y(z)$ has support on $[0,\infty)$ or on a nonnegative compact interval. Let $X\in\calX$ be a set of baseline covariates not affected by $Z$, and $U$ be a set of unmeasured baseline confounders also not affected by $Z$. Under the stable unit and treatment value assumption (SUTVA), the observed intermediate and final outcomes are $D = ZD(1) + (1-Z)D(0)$ and $Y = ZY(1) + (1-Z)Y(0)$, and the observed data vector is given by $O = (Z,D,Y,X)$. 

The principal stratification framework \citep{Frangakis2002} defines the principal strata by the joint values of potential intermediate outcomes under alternative treatment conditions, i.e., $S = (D(1), D(0)) \in \{11, 10, 01, 00\}$, corresponding to the pairs $(1,1), (1,0), (0,1)$, and $(0,0)$. It is instructive to emphasize that, within this framework, $S$ is conceptualized as an intrinsic latent characteristic of each unit, whereas $D(z)$ represents the realized partial manifestation of this characteristic under the treatment condition $Z = z$. Consequently, the observed treatment indicator $D$ is fully determined by the pair $(S, Z)$. Denote the marginal and conditional principal scores as $e_s = \bbP(S=s)$ and $e_s(x) = \bbP(S=s| X=x)$, respectively, for $s\in\{11,10,01,00\}$. We invoke the following structural assumptions.
\begin{assumption}[Treatment ignorability]\label{asp:treatment-ignorability}
    $Z \perp \{D(0), D(1), Y(0), Y(1)\} | X$.
\end{assumption}

\begin{assumption}[Latent principal ignorability] \label{asp:unconfoundedness-given-u}
    There exists a set of unmeasured baseline confounders $U$, not affected by $Z$, such that $Y(z) \perp S | X, U$. When $U \neq \emptyset$, this implies $Y(z) \not\perp S | X$, constituting a violation of the principal ignorability.
\end{assumption}

\begin{assumption}[Non-monotonicity] \label{asp:strata-identification}
    There exists a known function $\theta(\cdot): \calX\mapsto \Theta \equiv [0, \infty]$ such that $\theta(x) = e_{11}(x) e_{00}(x)/\{e_{10}(x) e_{01}(x)\}$.
\end{assumption}

Assumption \ref{asp:treatment-ignorability} assumes the standard treatment ignorability given $X$, which is automatically satisfied in randomized experiments and holds for observational studies where $X$ captures a sufficiently rich set of covariates that deconfound the $Z$--$Y$ relationship. Violations of this assumption in observational studies have been extensively investigated by \citet{Ding2014} and will therefore not be explored here. Assumption \ref{asp:unconfoundedness-given-u} is the structural relaxation of the principal ignorability (PI) assumption in \citet[Assumption 6]{Ding2016}, where $Y(z)$ and $S$ are conditionally independent given both $X$ and $U$, but not the measured covariate set $X$ itself. A slightly weaker form of the PI assumption is the mean principal ignorability (MPI) assumption, i.e., 
\begin{align*}
    \bbE\{Y(z) | D(0)=d^0, D(1)=d^1, X\} = \bbE\{Y(z) | D(z)=d^z, X\}, \quad z\in\{0,1\},
\end{align*}
which is also violated under Assumption \ref{asp:unconfoundedness-given-u} when $U\neq \emptyset$. (Detailed explanations are given in Section S2.1 of the Supplementary Materials.) Figure \ref{fig:DAG} provides an illustrative directed acyclic graph depicting the relationship among variables under Assumptions \ref{asp:treatment-ignorability} and \ref{asp:unconfoundedness-given-u}.

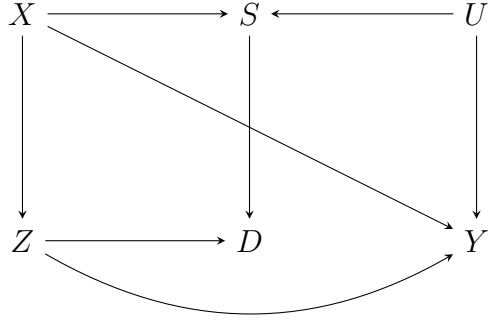
\begin{figure}[htbp]
    \centering
    \begin{tikzpicture}[>=stealth, node distance=2cm]
        \node (Z) at (0,0) {$Z$};
        \node (D) at (3,0) {$D$};
        \node (Y) at (6,0) {$Y$};
        \node (S) at (3,3) {$S$};
        \node (U) at (6,3) {$U$};
        \node (X) at (0,3) {$X$};
    
        \draw[->] (Z) -- (D);
        \draw[->] (Z) to[bend right=30] (Y);
        \draw[->] (S) -- (D);
        
        \draw[->] (U) -- (S);
        \draw[->] (U) -- (Y);
        
        \draw[->] (X) -- (Z);
        \draw[->] (X) -- (S);
        \draw[->] (X) -- (Y);
    \end{tikzpicture}
    \caption{A directed acyclic graph depicting the relationship among variables for principal stratification subject to unmeasured $S$--$Y$ confounding $U$.}
    \label{fig:DAG}
\end{figure}

Finally, in our setup, we follow \citet{Tong2025} to study the more general case of non-monotonicity, through an additional margin-free sensitivity parameter that quantifies the cross-world association between $D(1)$ and $D(0)$. Under the SUTVA and Assumption \ref{asp:treatment-ignorability}, the observed data only point identifies the marginal distributions of $D(z)$, i.e., $p_1(x)=\bbP(D=1 | Z=1, X=x) = e_{11}(x) + e_{10}(x)$ and $p_0(x)=\bbP(D=1 |Z=0, X=x) = e_{11}(x) + e_{01}(x)$. The system for $e_s(x)$ is undetermined even with the constraint that $e_{11}(x)+e_{10}(x)+e_{01}(x)+e_{00}(x)=1$. Under Assumptions \ref{asp:treatment-ignorability} and \ref{asp:strata-identification}, $\{e_{11}(x),e_{10}(x),e_{01}(x),e_{00}(x)\}$ are point identified as deterministic functions of $\{p_1(x), p_0(x), \theta(x)\}$. We provide the complete identification formulas for the principal scores in Section S2.2 of the Supplementary Materials.

Assumption \ref{asp:strata-identification} covers important special cases, and hence facilitates the generality of our key results. First, as the odds ratio parameter $\theta(x) \to \infty$, the population exhibits maximum positive dependence by forcing the denominator $e_{10}(x) e_{01}(x) \to 0$, which yields monotonicity. That is, $D(1) \geq D(0)$ almost surely when $p_1(x) > p_0(x)$, or $D(1) \leq D(0)$ almost surely when $p_1(x) < p_0(x)$. Conversely, setting $\theta(x) = 0$ drives the population to the lower bound of maximum negative dependence by forcing the numerator $e_{11}(x) e_{00}(x) = 0$, manifesting as either mutually exclusive compliance ($D(1)$ and $D(0)$ cannot simultaneously equal 1, i.e., $\bbP\{D(1) = 1, D(0) = 1 | X = x\} = 0$) when $p_1(x) + p_0(x) < 1$, or universal compliance ($D(1)$ and $D(0)$ cannot simultaneously equal 0, i.e., $\bbP\{D(1) = 0, D(0) = 0 | X = x\} = 0$) when $p_1(x) + p_0(x) > 1$. Between these structural extremes, the assumption of counterfactual independence occurs at $\theta(x) = 1$ \citep{hayden2005estimator}, where $D(1) \perp D(0) | X$ and there are still four principal strata (and hence non-monotonicity holds).

\section{Sensitivity bounds on the PCEs} \label{sec:PCE-bounds}

\subsection{Sensitivity parameters}

The observed data can be divided into cells in the form of $(Z=z, D=d)$ for $z,d\in\{0,1\}$, where each cell contains a permissible strata set $\calS_{z,d}$. Specifically, $\calS_{1,1} = \{10, 11\}$, $\calS_{0,0} = \{10, 00\}$, $\calS_{1,0} = \{00, 01\}$, and $\calS_{0,1} = \{11, 01\}$. For $z,d\in\{0,1\}$, we define $\RR_{SU}^{z,d}(x)\in[1,\infty)$ as the maximum relative risk (RR) representing the imbalance in the prevalence of the unmeasured confounder $U$ between the target stratum $s$ and the nuisance stratum $s'$ within the covariate stratum $X=x$, i.e.,
\begin{align} \label{eq:RR-SU}
    \RR_{SU}^{z,d}(x) = \max_u \frac{\bbP(U=u |S=s, X=x)}{\bbP(U=u |S=s', X=x)},
\end{align}
where $s, s' \in \calS_{z,d}$ with $s\neq s'$. We label $\RR_{SU}^{z,d}(x)$ as the selection RR. For ease of notation, we express $U$ as a categorical variable. If $U$ is continuous or mixed, $\bbP(U=u|\cdot)$ in \eqref{eq:RR-SU} is replaced by the corresponding probability density or general probability measure, and all subsequent bounding results hold via the Radon-Nikodym derivative, following the measure-theoretic extensions in \citet[eAppendix]{Ding2016epi}. Along this line, we also define $\RR_{UY}^z(x)\in[1,\infty)$ as the maximum RR of the unmeasured confounder $U$ on $Y(z)$ such that
\begin{align} \label{eq:RR-UY}
    \RR_{UY}^z(x) = \frac{\max_u \bbE\{Y(z) |X=x, U=u\}}{\min_u \bbE\{Y(z) |X=x, U=u\}},
\end{align}
We label $\RR_{UY}^z(x)$ as the outcome RR, which is independent of $S$ by Assumption \ref{asp:unconfoundedness-given-u}.

The RRs in \eqref{eq:RR-SU} and \eqref{eq:RR-UY} generalize the definitions in \citet{Ding2016epi} and \citet{Ding2016bmk} to principal stratification. Here, $\RR_{SU}^{z,d}(x)$ is the stratum-confounder imbalance parameter, which captures the $U \to S$ edge in Figure \ref{fig:DAG}. It quantifies how unevenly the unmeasured confounder $U$ is distributed between the target stratum $s$ and the nuisance stratum $s'$, answering the question of {\it ``How biologically or psychologically different are the two types of patients in the cell $(Z=z,D=d)$?''} Whereas $\RR_{UY}^z(x)$ is the confounder-outcome parameter, which captures the structural $U \to Y$ edge in Figure \ref{fig:DAG}. It quantifies the maximum impact the unmeasured confounder $U$ has on the expected potential outcome, answering the question of: {\it ``How severely does the unmeasured confounder alter the latent disease trajectory $Y(z)$ for any patient regardless of which treatment cell they ultimately occupy?''} Both RRs are margin-free, as they do not depend on units or scales of $U$ and $Y(z)$. 

We then define the latent stratum mean $\mu_{z,s}(x)=\bbE\{Y(z)|S=s,X=x\}=\bbE(Y|Z=z,S=s,X=x)$ and the observed cell mean $m_{z,d}(x)=\bbE(Y|Z=z,D=d,X=x)$. The MPI assumption implies that $\mu_{z,s}(x)=m_{z,d}(x)$ for all $s\in\calS_{z,d}$. The sensitivity parameters $\RR_{SU}^{z,d}(x)$ and $\RR_{UY}^z(x)$ are active when the presence of $U$ forces $\mu_{z,s}(x)$ and $\mu_{z,s'}(x)$ to diverge from $m_{z,d}(x)$, thus violating the MPI assumption. Mathematically, the confounding by $U$ only exists if $\RR_{SU}^{z,d}(x) > 1$ and $\RR_{UY}^z(x) > 1$. If $\RR_{SU}^{z,d} = 1$, then $U$ is perfectly balanced between $s$ and $s'$, severing the $U \to S$ edge in Figure \ref{fig:DAG}. Similarly, if $\RR_{UY}^z(x) = 1$, meaning the maximum and minimum expectations of the outcome across all possible strata of $U$ are identical, then $U$ has no effect on the trajectory of $Y(z)$. Thus, the $U \to Y$ edge in Figure \ref{fig:DAG} is severed as well.

For any observed cell $(Z=z, D=d)$ composed of a target stratum $s$ and a nuisance stratum $s'$, we first define the cell-specific bounding factor $\BF_U^{z,d}(x)$ as the following deterministic function of the two RRs, i.e.,
\begin{align} \label{eq:BF}
    \BF_U^{z,d}(x) = \frac{\RR_{SU}^{z,d}(x)\RR_{UY}^z(x)}{\RR_{SU}^{z,d}(x) + \RR_{UY}^z(x) - 1}.
\end{align}
By definition, $\BF_U^{z,d}(x)\leq\min\{\RR_{SU}^{z,d}(x),\RR_{UY}^z(x)\}$, and $\BF_U^{z,d}(x)$ is also margin-free. We then have the following result regarding $\BF_U^{z,d}(x)$ in Lemma \ref{lem:BF}.
\begin{lemma} \label{lem:BF}
    Under Assumptions \ref{asp:treatment-ignorability} and \ref{asp:unconfoundedness-given-u}, the ratio of $\mu_{z,s}(x)$ to $\mu_{z,s'}(x)$ is bounded by $\BF_{U}^{z,d}(x)$ in \eqref{eq:BF} as
    \begin{align} \label{eq:ratio-bounds}
        \frac{1}{\BF_U^{z,d}(x)} \leq \frac{\mu_{z,s'}(x)}{\mu_{z,s}(x)} \leq \BF_U^{z,d}(x).
    \end{align}
\end{lemma}

Lemma \ref{lem:BF} extends \citet{Ding2016epi} to principal stratification with a binary intermediate variable. By definition, the RRs are bounded below by 1, which guarantees that $\BF_U^{z,d}(x) \geq 1$. This mathematical floor represents the limiting state of Assumption \ref{asp:unconfoundedness-given-u}. Specifically, if $U$ is either perfectly balanced across the latent strata ($\RR_{SU}^{z,d}(x) = 1$) or exerts zero effect on the outcome trajectory ($\RR_{UY}^z(x) = 1$), the bounding factor collapses to $\BF_U^{z,d}(x) = 1$. Under this limit, the bounding inequalities in \eqref{eq:ratio-bounds} perfectly tighten ($1 \leq \mu_{z,s'}(x) / \mu_{z,s}(x) \leq 1$), forcing $\mu_{z,s}(x) = \mu_{z,s'}(x)$. This algebraically recovers the MPI assumption, allowing the observable mixture mean to point identify the latent stratum means ($m_{z,d}(x) = \mu_{z,s}(x)$ for all $s\in\calS_{z,d}$). Thus, $\BF_U^{z,d}(x) > 1$ serves as the exact quantitative measure of MPI violation. Finally, it is worth stating that unlike the standard setting of \citet{Ding2016epi} and \citet{Ding2016bmk}, our target estimand $\mu_{z,s}(x)$ is latent and enters $m_{z,d}(x)$ as a weighted mixture with $\mu_{z,s'}(x)$, so the bounding factor must propagate through the mixture decomposition. The mixing weights $w_{z,s}(x)$ themselves also depend on $\theta(x)$ through Assumption \ref{asp:strata-identification}, adding an additional sensitivity dimension, which we will exploit in the subsequent bounds derivation.

\subsection{General bounds}

To proceed, we define the observed cell-specific mixing weights $w_{z,s}(x)$ for $s\in\calS_{z,d}$, where $w_{z,s}(x)=e_s(x)/p_z(x)$ for $d=1$ and $w_{z,s}(x)=e_s(x)/\{1-p_z(x)\}$ for $d=0$. Theorem \ref{thm:general-bounds} gives the sharp nonparametric sensitivity bounds on $\mu_{z,s}(x)$.

\begin{theorem} \label{thm:general-bounds}
    Under Assumptions \ref{asp:treatment-ignorability}--\ref{asp:strata-identification}, $\mu_{z,s}(x)$ is sharply bounded by $L_{z,s}(x)\leq \mu_{z,s}(x) \leq U_{z,s}(x)$, where
    \begin{align*} 
        L_{z,s}(x) &= \max\left[ \frac{m_{z,d}(x)}{w_{z,s}(x) + \{1 - w_{z,s}(x)\} \BF_U^{z,d}(x)}, 1 - \frac{1 - m_{z,d}(x)}{w_{z,s}(x) + \{1 - w_{z,s}(x)\} / \BF_U^{z,d}(x)} \right],\displaybreak[0]\\
        U_{z,s}(x) &= \min\left[ \frac{m_{z,d}(x)}{w_{z,s}(x) + \{1 - w_{z,s}(x)\} / \BF_U^{z,d}(x)}, 1 - \frac{1 - m_{z,d}(x)}{w_{z,s}(x) + \{1 - w_{z,s}(x)\} \BF_U^{z,d}(x)} \right],
    \end{align*}
    for $s\in\calS_{z,d}$.
\end{theorem}

Each bound in Theorem \ref{thm:general-bounds} contains two components: one from directly applying \eqref{eq:ratio-bounds}, and the other from applying \eqref{eq:ratio-bounds} to $1-Y(z)$ so that the obtained final bounds stay valid, since $Y(z)\in[0,1]$. To obtain the covariate-adjusted bounds on the marginal expectation $\mu_{z,s} \equiv \bbE\{\mu_{z,s}(x)|S=s\}$, we can directly integrate the conditional bounds in Theorem \ref{thm:general-bounds} over the stratum-specific covariate distribution $F_{X | S=s}(x)$. Next, Theorem \ref{thm:general-bounds} also implies the sharp nonparametric bounds on the conditional PCE $\Delta_s(x) \equiv \mu_{1,s}(x) - \mu_{0,s}(x)$, given in Corollary \ref{coro:PCE-bounds}

\begin{corollary} \label{coro:PCE-bounds}
    Under Assumptions \ref{asp:treatment-ignorability}--\ref{asp:strata-identification}, $\Delta_s(x)$ is sharply bounded by 
    \begin{align*}
        L_{1,s}(x) - U_{0,s}(x) \leq \Delta_s(x) \leq U_{1,s}(x) - L_{0,s}(x).
    \end{align*}
    To obtain bounds on the marginal PCE $\Delta_s\equiv\bbE\{\Delta_s(x)|S=s\}$, the conditional bounds can be integrated over the stratum-specific covariate distribution $F_{X | S=s}(x)$.   
\end{corollary}

Interestingly, although our bounds are motivated by the violation of MPI, they have connections to an existing bound derived in the principal stratification literature. For example, let $[L_{z,s}^\mathrm{GM}(x), U_{z,s}^\mathrm{GM}(x)]$ denote the conditional nonparametric bounds on $\mu_{z,s}(x)$ introduced by \citet{Grilli2008}, which were originally established in \citet{Zhang2003}, where
\begin{align} \label{eq:GM-bounds}
    L_{z,s}^\mathrm{GM}(x) = \max\left\{ 0, 1 - \frac{1 - m_{z,d}(x)}{w_{z,s}(x)} \right\} \quad\text{and}\quad U_{z,s}^\mathrm{GM}(x) = \min\left\{ 1, \frac{m_{z,d}(x)}{w_{z,s}(x)} \right\}.
\end{align}
Although the bounds from \citet{Grilli2008} were initially derived in a setting without covariates, their formulation can be straightforwardly generalized to accommodate covariate information as in \eqref{eq:GM-bounds}. Proposition \ref{prop:GM-bounds} presents results regarding the relationship between $[L_{z,s}(x),U_{z,s}(x)]$ and $[L_{z,s}^\mathrm{GM}(x), U_{z,s}^\mathrm{GM}(x)]$.

\begin{proposition} \label{prop:GM-bounds}
    Under Assumptions \ref{asp:treatment-ignorability}--\ref{asp:strata-identification}, for any $\BF_U^{z,d}(x)\in(1,\infty)$, we have $L_{z,s}^\mathrm{GM}(x) < L_{z,s}(x) \leq U_{z,s}(x) < U_{z,s}^\mathrm{GM}(x)$. As $\BF_U^{z,d}(x) \to \infty$, $L_{z,s}(x) \downarrow L_{z,s}^\mathrm{GM}(x)$ and $U_{z,s}(x) \uparrow U_{z,s}^\mathrm{GM}(x)$.
\end{proposition}

Proposition \ref{prop:GM-bounds} proves that for any finite $\BF_U^{z,d}(x)$, the proposed bounds nest within those of \citet{Grilli2008}, and as $\BF_U^{z,d}(x)$ approaches infinity, the proposed bounds converge to those of \citet{Grilli2008}. This result is intuitive, as the bounds of \citet{Grilli2008} were established in the worst-case scenario, where the unmeasured confounding is the strongest, characterized by $\BF_U^{z,d}(x)\to\infty$. This relationship further indicates that the bounding factor $\BF_U^{z,d}(x)$ functions as a continuous sensitivity bridge, providing a smooth continuum between the worst-case and alternative settings with weaker degrees of unmeasured confounding.

\subsection{Bounds under monotonicity}

We next consider the special case where $\theta(x) \to \infty$ such that the monotonicity assumption holds with $D(1) \geq D(0)$ (e.g., one-sided non-compliance), implying $e_{01}(x) = 0$. This leads to $\calS_{1,1} = \{10, 11\}$, $\calS_{0,0} = \{10, 00\}$, $\calS_{1,0} = \{00\}$, and $\calS_{0,1} = \{11\}$, with the point identification of $\mu_{0,11}(x)=m_{0,1}(x)$ and $\mu_{1,00}(x)=m_{1,0}(x)$. Therefore, for the point identification sets $\calS_{1,0}$ and $\calS_{0,1}$, $L_{0,11}(x)=U_{0,11}(x)=m_{0,1}(x)$ and $L_{1,00}(x)=U_{1,00}(x)=m_{1,0}(x)$. The RRs in \eqref{eq:RR-SU} and \eqref{eq:RR-UY} are only meaningfully defined for $\calS_{1,1}$ and $\calS_{0,0}$. The conditional principal scores for the remaining strata are $e_{11}(x) = p_0(x)$, $e_{10}(x) = p_1(x) - p_0(x)$, and $e_{00}(x) = 1 - p_1(x)$. The mixing weights are $w_{1,s}(x)=e_s(x)/p_s$ for $s\in\calS_{1,1}$ and $w_{0,s}(x)=e_s(x)/p_s$ for $s\in\calS_{0,0}$. Theorem \ref{thm:general-bounds} implies the bounds under monotonicity given in Corollary \ref{coro:monotone-bounds}.

\begin{corollary} \label{coro:monotone-bounds}
    Under Assumptions \ref{asp:treatment-ignorability}--\ref{asp:strata-identification} with $\theta(x)\to\infty$, $\mu_{z,s}(x)$ is sharply bounded by $L_{z,s}(x)\leq \mu_{z,s}(x) \leq U_{z,s}(x)$, where
    \begin{align*} 
        L_{z,s}(x) &= \max\left[ \frac{m_{z,d}(x)}{w_{z,s}(x) + \{1 - w_{z,s}(x)\} \BF_U^{z,d}(x)}, 1 - \frac{1 - m_{z,d}(x)}{w_{z,s}(x) + \{1 - w_{z,s}(x)\} / \BF_U^{z,d}(x)} \right], \displaybreak[0]\\
        U_{z,s}(x) &= \min\left[ \frac{m_{z,d}(x)}{w_{z,s}(x) + \{1 - w_{z,s}(x)\} / \BF_U^{z,d}(x)}, 1 - \frac{1 - m_{z,d}(x)}{w_{z,s}(x) + \{1 - w_{z,s}(x)\} \BF_U^{z,d}(x)} \right],
    \end{align*}
    for $s\in\calS_{1,1},\calS_{0,0}$. For $s\in\calS_{1,0},\calS_{0,1}$, $L_{0,11}(x)=U_{0,11}(x)=m_{0,1}(x)$ and $L_{1,00}(x)=U_{1,00}(x)=m_{1,0}(x)$.
\end{corollary}

Let $[L_{z,s}^\mathrm{LH}(x), U_{z,s}^\mathrm{LH}(x)]$ denote the conditional worst-case nonparametric bounds on $\mu_{z,s}(x)$ introduced by \citet{Long2013} under the monotonicity assumption, where $[L_{z,s}^\mathrm{LH}(x), U_{z,s}^\mathrm{LH}(x)]$ has the same form as $[L_{z,s}^\mathrm{GM}(x), U_{z,s}^\mathrm{GM}(x)]$, but only for $s\in\calS_{1,1},\calS_{0,0}$. Corollary \ref{prop:LH-bounds} presents results regarding the relationship between $[L_{z,s}(x),U_{z,s}(x)]$ and $[L_{z,s}^\mathrm{LH}(x), U_{z,s}^\mathrm{LH}(x)]$.

\begin{corollary} \label{prop:LH-bounds}
    Under Assumptions \ref{asp:treatment-ignorability}--\ref{asp:strata-identification} with $\theta(x)\to\infty$, for any $\BF_U^{z,d}(x) \in (1,\infty)$, we have $L_{z,s}^\mathrm{LH}(x) < L_{z,s}(x) \leq U_{z,s}(x) < U_{z,s}^\mathrm{LH}(x)$. As $\BF_U^{z,d}(x) \to \infty$, $L_{z,s}(x) \downarrow L_{z,s}^\mathrm{LH}(x)$ and $U_{z,s}(x) \uparrow U_{z,s}^\mathrm{LH}(x)$.
\end{corollary}

Corollary \ref{prop:LH-bounds} is a special case of Proposition \ref{prop:GM-bounds} under monotonicity. It states that, when monotonicity holds, for any finite $\BF_U^{z,d}(x)$, the proposed bounds nest within those of \citet{Long2013}, and as $\BF_U^{z,d}(x)$ approaches infinity, the proposed bounds converge to those of \citet{Long2013}. 

\subsection{Cornfield connection}

To facilitate the application of the derived bounds for sensitivity analysis, we establish a connection between the proposed bounds and the Cornfield-type conditions \citep{Cornfield2009} in the current context of principal stratification. Denote $L_{z,s}(x, \BF_U^{z,d})\equiv L_{z,s}(x)$ and $U_{z,s}(x, \BF_U^{z,d})\equiv U_{z,s}(x)$. Without loss of generality, assume an apparently causal effect, where the point estimate of $\Delta_s(x)$ under the MPI assumption is positive. To determine the falsification boundary, we define the maximum bounding factor across treatment arms as $\Lambda_s(x) \equiv \max_{z \in \{0,1\}} \BF_U^{z, d}(x)$, where $s\in\calS_{z,d}$. Provided the worst-case lower bound crosses the null ($\lim_{\Lambda_s(x) \to \infty} \{ L_{1,s}(x,\Lambda_s) - U_{0,s}(x,\Lambda_s) \} \leq 0$), the critical falsification threshold $\Lambda_s^*(x)$ is point identified as the unique real root of the piecewise equation $L_{1,s}(x,\Lambda_s) - U_{0,s}(x,\Lambda_s) = 0$ on the interval $(1, \infty)$. Here, $L_{1,s}(x,\Lambda_s)$ and $U_{0,s}(x,\Lambda_s)$ respectively denote the bounds $L_{1,s}(x, \BF_U^{1,d})$ and $U_{0,s}(x, \BF_U^{0,d})$, where all constituent bounding factors are replaced by $\Lambda_s(x)$. Theorem \ref{thm:cornfield} provides the necessary conditions for the RRs to attain this boundary.

\begin{theorem} \label{thm:cornfield}
    Under Assumptions \ref{asp:treatment-ignorability}--\ref{asp:strata-identification}, to generate a confounding magnitude sufficient to nullify $\Delta_s(x)$, the unmeasured confounder $U$ must satisfy the following Cornfield-type conditions for the RRs of at least one of the treatment arms $z \in \{0, 1\}$:
    \begin{align}
        \min\left\{\RR_{SU}^{z,d}(x), \RR_{UY}^z(x)\right\} &\geq \Lambda_s^*(x), \label{eq:cornfield-condition-1}\displaybreak[0]\\
        \max\left\{\RR_{SU}^{z,d}(x), \RR_{UY}^z(x)\right\} &\geq \Lambda_s^*(x) + \sqrt{\Lambda_s^*(x)\{\Lambda_s^*(x) - 1\}}. \label{eq:cornfield-condition-2}
    \end{align}
\end{theorem}

Theorem \ref{thm:cornfield} establishes the joint magnitude of unmeasured confounding required to nullify the PCE in stratum $s$, $\Delta_s(x)$. The low threshold in \eqref{eq:cornfield-condition-1} formalizes the condition that a confounder cannot nullify the effect by operating exclusively through a single mechanism (either selection or outcome RR); it must simultaneously exert an RR of at least $\Lambda_s^*(x)$ on both principal stratum membership and the potential outcome. A weak association in one pathway cannot be offset by an infinitely strong association in the other. Furthermore, the high threshold in \eqref{eq:cornfield-condition-2} dictates that satisfying the bare minimum across both pathways is mathematically insufficient. To drive the lower bound to zero, the dominant confounding pathway must clear a higher boundary. Importantly, the high threshold in \eqref{eq:cornfield-condition-2} also motivates the \emph{principal E-value} in the context of principal stratification. That is, given an estimated $\Delta_s(x) > 0$, the \emph{principal E-value} within stratum $S=s$ is defined as
\begin{align} \label{eq:e-value}
    \mathrm{EV}_s(x) \equiv \Lambda_s^*(x) + \sqrt{\Lambda_s^*(x)\{\Lambda_s^*(x) - 1\}},
\end{align}
which is the minimum risk ratio that the unmeasured confounder $U$ must share with both principal stratum membership (selection RR) and the potential outcome (outcome RR) in at least one treatment arm $z \in \{0,1\}$ to fully explain away the PCE $\Delta_s(x)$.

\subsection{Bounds for binary outcome}

Finally, we address the special case with a ratio effect measure when $Y(z)\in\{0,1\}$. Besides the difference scale estimand $\Delta_s(x)$, for binary outcomes, define the multiplicative estimands, the conditional principal causal risk ratio (PCRR) and odds ratio (PCOR), respectively, as
\begin{align} \label{eq:pcrr-pcor-def}
    \mathrm{PCRR}_s(x) = \frac{\mu_{1,s}(x)}{\mu_{0,s}(x)} \quad\text{and}\quad \mathrm{PCOR}_s(x) = \frac{\mu_{1,s}(x) / \{1 - \mu_{1,s}(x)\}}{\mu_{0,s}(x) / \{1 - \mu_{0,s}(x)\}}.
\end{align}

Corollary \ref{coro:pcrr-pcor-bounds} gives the sharp nonparametric bounds on $\mathrm{PCRR}_s(x)$ and $\mathrm{PCOR}_s(x)$.

\begin{corollary} \label{coro:pcrr-pcor-bounds}
    Under Assumptions \ref{asp:treatment-ignorability}--\ref{asp:strata-identification}, the $\mathrm{PCRR}_s(x)$ and $\mathrm{PCOR}_s(x)$ in \eqref{eq:pcrr-pcor-def} are sharply bounded by
    \begin{align*}
        \frac{L_{1,s}(x)}{U_{0,s}(x)} \leq \mathrm{PCRR}_s(x) \leq \frac{U_{1,s}(x)}{L_{0,s}(x)}
    \end{align*}
    and
    \begin{align*}
        \frac{L_{1,s}(x) / \{1 - L_{1,s}(x)\}}{U_{0,s}(x) / \{1 - U_{0,s}(x)\}} \leq \mathrm{PCOR}_s(x) \leq \frac{U_{1,s}(x) / \{1 - U_{1,s}(x)\}}{L_{0,s}(x) / \{1 - L_{0,s}(x)\}},
    \end{align*}
    respectively.
\end{corollary}

Let $\Lambda_s^*(x)$ be the unique critical falsification threshold derived in Theorem \ref{thm:cornfield} required to nullify $\Delta_s(x)$, which is the conditional principal causal risk difference when $Y(z)$ is binary, such that $L_{1,s}(x, \Lambda_s^*) = U_{0,s}(x, \Lambda_s^*)$. Then, we have the following results in Corollary \ref{coro:E-value}, which states that the Cornfield-type thresholds for $\RR_{SU}^{z,d}(x)$ and $\RR_{UY}^z(x)$ in Theorem \ref{thm:cornfield} are invariant across causal contrast scales.

\begin{corollary} \label{coro:E-value}
    Under Assumptions \ref{asp:treatment-ignorability}--\ref{asp:strata-identification}, the critical falsification threshold $\Lambda_s^*(x)$ given in Theorem \ref{thm:cornfield} simultaneously nullifies $\mathrm{PCRR}_s(x)$ and $\mathrm{PCOR}_s(x)$. Specifically, the condition $L_{1,s}(x, \Lambda_s^*) = U_{0,s}(x, \Lambda_s^*)$ ensures that $L_{1,s}(x, \Lambda_s^*)/U_{0,s}(x, \Lambda_s^*) = 1$ and $[L_{1,s}(x, \Lambda_s^*) / \{1 - L_{1,s}(x, \Lambda_s^*)\}]/[U_{0,s}(x, \Lambda_s^*) / \{1 - U_{0,s}(x, \Lambda_s^*)\}] = 1$.
\end{corollary}

\section{Extension to the pairwise comparison estimands} \label{sec:PGCE-bounds}

\subsection{The PGCE estimand}

As an estimand, the PCE has been predominantly formulated for continuous and binary outcomes. This focus limits its applicability to causal inference in settings with more complex outcome types, such as categorical, ordinal, or multiple responses of mixed types. To allow for general causal comparisons across more outcome-types in the presence of intermediate variables, \citet{Chen2026} introduced the principal generalized causal effect (PGCE) estimand:
\begin{align} \label{eq:def-pce}
    \mu_{z_1,z_2,s}=\bbE[h\{Y_1(z_1),Y_2(z_2)\}|S_1=S_2=s],
\end{align}
for $s\in\{11,10,01,00\}$ and $z_1\neq z_2$, where $h(u,v)$ is a general contrast function defined on the product space $\calY\times\calY$, e.g., $h(u,v) = \bbone(u\geq v)$ and $h(u,v)=u-v$, which recovers the PCE. The outcome space $\calY$ can be relaxed if the range of $h$ is bounded, e.g., $h(u,v)=\bbone(u\geq v)\in[0,1]$. The conditional PGCE is $\mu_{z_1,z_2,s}(x_1,x_2)\equiv \bbE[h\{Y_1(z_1),Y_2(z_2)\}|S_1=S_2=s,X_1=x_1,X_2=x_2]$ with $\mu_{z_1,z_2,s}=\bbE\{\mu_{z_1,z_2,s}(x_1,x_2)|S_1=S_2=s\}$. Here, $\{Y_1(z_1),Z_1,S_1,X_1\}$ and $\{Y_2(z_2),Z_2,S_2,X_2\}$ are independent replicates from the same data-generating process. The definition in \eqref{eq:def-pce} responds to the fact that the alternative estimand $\bbE[h\{Y(z),Y(1-z)\}|S=s]$ is generally inestimable for nonlinear $h(u,v)$ because $Y(0)$ and $Y(1)$ are never simultaneously observed, and is motivated from the literature on pairwise comparison estimands \citep{Mao2018,mao2024wilcoxon}. 

For the PGCE estimand, the MPI assumption for the point identification of $\mu_{z_1,z_2,s}(x_1,x_2)$ becomes
\begin{align*}
    &\bbE[h\{Y_1(z_1),Y_2(z_2)\}|D_1(0)=d_1^0,D_1(1)=d_1^1,D_2(0)=d_2^0,D_2(1)=d_2^1,X_1=x_1,X_2=x_2]\displaybreak[0]\\
    &=\bbE[h\{Y_1(z_1),Y_2(z_2)\}|D_1(z_1)=d_1^{z_1},D_2(z_2)=d_2^{z_2},X_1=x_1,X_2=x_2],
\end{align*}
which is violated under Assumption \ref{asp:unconfoundedness-given-u}. Quantities in the form of $\mu_{z_1,z_2,s_1,s_2} \equiv \bbE[h\{Y_1(z_1),Y_2(z_2)\}|S_1=s_1,S_2=s_2]$ with $s_1 \neq s_2$ are point identifiable under Assumptions \ref{asp:treatment-ignorability} and \ref{asp:strata-identification}, together with the MPI. However, as discussed in \citet{Chen2026}, they lack a meaningful causal interpretation due to comparing potential outcomes from two distinct strata. Thus, we restrict attention to $\mu_{z_1,z_2,s}$ with $s_1 = s_2 = s$, and provide the following example to provide a concrete context. 

\begin{example}
    An example of $\mu_{z_1,z_2,s}$ is the probabilistic index under truncation by death. Specifically, when $D$ represents the survival status (alive if $D=1$), $s=11$ represents the always-survivors who would survive until the measurement of $Y$ regardless of $z$; $s=10$ and $s=01$ represent the protected and harmed, who would survive only under $z=1$ and $z=0$, respectively; $s=00$ represents the never-survivors who would not survive regardless of $z$. In clinical research, \citet{Acion2006} argued that an intuitive effect size can be measured by the probabilistic index statistic defined with $h(u,v)=\bbone(u\geq v)$, leading to the survivor probabilistic index as $\mu_{1,0,11}=\bbP\{Y_1(1)\geq Y_2(0)|S_1=S_2=11\}$.
\end{example}

Before we proceed, we emphasize that the extension to PGCE estimands with nonlinear contrast functions $h$ presents new challenges beyond a product-space application of Section~\ref{sec:PCE-bounds}. When $h$ is nonlinear, the two potential outcomes enter non-separably, so the outcome relative risk can no longer be defined marginally for each unit; it must instead be defined jointly over the unmeasured confounders of both units simultaneously. This non-separability propagates through the entire bounding framework, producing a richer sensitivity parameterization and a more complex mixture structure.

\subsection{Sensitivity parameters}

For PGCEs with linear contrast functions, e.g., $h(u,v)=u-v$, the bounds can be directly obtained following Section \ref{sec:PCE-bounds}. Therefore, here, we focus on deriving bounds on the PGCEs with nonlinear contrast functions such that $h:\calY\times\calY\mapsto[0,1]$. Since $h\{Y_1(z_1),Y_2(z_2)\}$ is non-separable, we define $\RR_{UY}^{z_1,z_2}(x_1,x_2)\in[1,\infty)$ analogous to $\RR_{UY}^z(x)$ in \eqref{eq:RR-UY} as the maximum RRs of the unmeasured confounders $U_1$ and $U_2$ on $h\{Y_1(z_1),Y_2(z_2)\}$ such that
\begin{align} 
    \RR_{UY}^{z_1,z_2}(x_1,x_2) = \frac{\max_{u_1,u_2} \bbE[h\{Y_1(z_1),Y_2(z_2)\} |X_j=x_j, U_j=u_j, j=1,2\}}{\min_{u_1,u_2} \bbE[h\{Y_1(z_1),Y_2(z_2)\} |X_j=x_j, U_j=u_j, j=1,2\}}, \label{eq:PGCE-RR-UY}
\end{align}

For observed cells pair $(Z_1=z_1, D_1=d_1)$ and $(Z_2=z_2, D_2=d_2)$ composed of a target stratum $s$ and nuisance strata $s_1'$ and $s_2'$, respectively, we define the cell-pair-specific bounding factors $\BF_{U[1]}^{z_1,z_2,d_1,d_2}(x_1,x_2)$, $\BF_{U[2]}^{z_1,z_2,d_1,d_2}(x_1,x_2)$, and $\BF_{U[1,2]}^{z_1,z_2,d_1,d_2}(x_1,x_2)$ as the following deterministic functions of the $\RR_{SU}^{z,d}(x)$ in \eqref{eq:RR-SU} and the outcome RR in \eqref{eq:PGCE-RR-UY}, i.e.,
\begin{align} 
    \BF_{U[1]}^{z_1,z_2,d_1,d_2}(x_1,x_2) &= \frac{\RR_{SU}^{z_1,d_1}(x_1) \RR_{UY}^{z_1,z_2}(x_1,x_2)}{\RR_{SU}^{z_1,d_1}(x_1) + \RR_{UY}^{z_1,z_2}(x_1,x_2) - 1}, \label{eq:PGCE-BF-1} \displaybreak[0]\\
    \BF_{U[2]}^{z_1,z_2,d_1,d_2}(x_1,x_2) &= \frac{\RR_{SU}^{z_2,d_2}(x_2) \RR_{UY}^{z_1,z_2}(x_1,x_2)}{\RR_{SU}^{z_2,d_2}(x_2) + \RR_{UY}^{z_1,z_2}(x_1,x_2) - 1}, \label{eq:PGCE-BF-2} \displaybreak[0]\\
    \BF_{U[1,2]}^{z_1,z_2,d_1,d_2}(x_1,x_2) &= \frac{\RR_{SU}^{z_1,d_1}(x_1)\RR_{SU}^{z_2,d_2}(x_2) \RR_{UY}^{z_1,z_2}(x_1,x_2)}{\RR_{SU}^{z_1,d_1}(x_1)\RR_{SU}^{z_2,d_2}(x_2) + \RR_{UY}^{z_1,z_2}(x_1,x_2) - 1}. \label{eq:PGCE-BF-1-2}
\end{align}
Since $\RR_{SU}^{z,d}(x)\geq 1$, we can verify the following ordering relationship: 
\begin{align} \label{eq:PGCE-BF-relationship}
    \BF_{U[1,2]}^{z_1,z_2,d_1,d_2}(x_1,x_2) \geq \max\left\{\BF_{U[1]}^{z_1,z_2,d_1,d_2}(x_1,x_2),\BF_{U[2]}^{z_1,z_2,d_1,d_2}(x_1,x_2)\right\}.
\end{align}

Let $\mu_{z_1,z_2,s_1,s_2}(x_1,x_2) = \bbE[h\{Y_1(z_1),Y_2(z_2)\}|S_1=s_1,S_2=s_2,X_1=x_1,X_2=x_2]$ with $s_1 \neq s_2$. We first obtain the following result regarding \eqref{eq:PGCE-BF-1}--\eqref{eq:PGCE-BF-1-2} in Lemma \ref{lem:PGCE-BF}.
\begin{lemma} \label{lem:PGCE-BF}
    Under Assumptions \ref{asp:treatment-ignorability} and \ref{asp:unconfoundedness-given-u}, we have
    \begin{align} 
        \frac{1}{\BF_{U[1]}^{z_1,z_2,d_1,d_2}(x_1,x_2)} &\leq \frac{\mu_{z_1,z_2,s_1',s}(x_1,x_2)}{\mu_{z_1,z_2,s}(x_1,x_2)} \leq \BF_{U[1]}^{z_1,z_2,d_1,d_2}(x_1,x_2), \label{eq:PGCE-ratio-bounds-1} \displaybreak[0]\\
        \frac{1}{\BF_{U[2]}^{z_1,z_2,d_1,d_2}(x_1,x_2)} &\leq \frac{\mu_{z_1,z_2,s,s_2'}(x_1,x_2)}{\mu_{z_1,z_2,s}(x_1,x_2)} \leq \BF_{U[2]}^{z_1,z_2,d_1,d_2}(x_1,x_2), \label{eq:PGCE-ratio-bounds-2} \displaybreak[0]\\
        \frac{1}{\BF_{U[1,2]}^{z_1,z_2,d_1,d_2}(x_1,x_2)} &\leq \frac{\mu_{z_1,z_2,s_1',s_2'}(x_1,x_2)}{\mu_{z_1,z_2,s}(x_1,x_2)} \leq \BF_{U[1,2]}^{z_1,z_2,d_1,d_2}(x_1,x_2). \label{eq:PGCE-ratio-bounds-1-2}
    \end{align}
\end{lemma}

Lemma \ref{lem:PGCE-BF} generalized Lemma \ref{lem:BF} to the pairwise comparison context, where \eqref{eq:PGCE-ratio-bounds-1} and \eqref{eq:PGCE-ratio-bounds-2} respectively bounds the ratio of $\mu_{z_1,z_2,s_1,s}(x_1,x_2)$ and $\mu_{z_1,z_2,s,s_2}(x_1,x_2)$ to $\mu_{z_1,z_2,s}(x_1,x_2)$ with one latent stratum being different from $s$, and \eqref{eq:PGCE-ratio-bounds-1-2} bounds the ratio of $\mu_{z_1,z_2,s_1,s_2}(x_1,x_2)$ to $\mu_{z_1,z_2,s}(x_1,x_2)$ with both latent strata being different from $s$.

\subsection{General bounds}

With Lemma \ref{lem:PGCE-BF}, the sharp nonparametric bounds on $\mu_{z_1,z_2,s}(x_1,x_2)$ can be derived, and their explicit forms are given in Theorem \ref{thm:PGCE-general-bounds}.

\begin{theorem} \label{thm:PGCE-general-bounds}
    Under Assumptions \ref{asp:treatment-ignorability}--\ref{asp:strata-identification}, for $s\in\calS_{z_1,d_1}\bigcap\calS_{z_2,d_2}$, $\mu_{z_1,z_2,s}(x_1,x_2)$ is sharply bounded by $L_{z_1,z_2,s}(x_1,x_2)\leq \mu_{z_1,z_2,s}(x_1,x_2) \leq U_{z_1,z_2,s}(x_1,x_2)$, where
    \begin{align*}
        &L_{z_1,z_2,s}(x_1,x_2) \displaybreak[0]\\
        &= \max\Bigg( m_{z_1,z_2,d_1,d_2}(x_1,x_2) \Big[ w_{z_1,s}(x_1)w_{z_2,s}(x_2) + \{1 - w_{z_1,s}(x_1)\}w_{z_2,s}(x_2) \BF_{U[1]}^{z_1,z_2,d_1,d_2}(x_1,x_2) \displaybreak[0]\\
        &\qquad\qquad\qquad\qquad\qquad\qquad + w_{z_1,s}(x_1)\{1 - w_{z_2,s}(x_2)\} \BF_{U[2]}^{z_1,z_2,d_1,d_2}(x_1,x_2) \displaybreak[0]\\
        &\qquad\qquad\qquad\qquad\qquad\qquad + \{1 - w_{z_1,s}(x_1)\}\{1 - w_{z_2,s}(x_2)\} \BF_{U[1,2]}^{z_1,z_2,d_1,d_2}(x_1,x_2)\Big]^{-1}, \displaybreak[0]\\
        &\qquad\qquad 1- \{1-m_{z_1,z_2,d_1,d_2}(x_1,x_2)\} \displaybreak[0]\\
        &\qquad\qquad\qquad\times\Bigg[ w_{z_1,s}(x_1)w_{z_2,s}(x_2) + \frac{\{1 - w_{z_1,s}(x_1)\}w_{z_2,s}(x_2)}{\BF_{U[1]}^{z_1,z_2,d_1,d_2}(x_1,x_2)} \displaybreak[0]\\
        &\qquad\qquad\qquad\qquad + \frac{w_{z_1,s}(x_1)\{1-w_{z_2,s}(x_2)\}}{\BF_{U[2]}^{z_1,z_2,d_1,d_2}(x_1,x_2)} + \frac{\{1 - w_{z_1,s}(x_1)\}\{1-w_{z_2,s}(x_2)\}}{\BF_{U[1,2]}^{z_1,z_2,d_1,d_2}(x_1,x_2)} \Bigg]^{-1} \Bigg)
    \end{align*}
    and
    \begin{align*}
        &U_{z_1,z_2,s}(x_1,x_2) \displaybreak[0]\\
        &= \min\Bigg( m_{z_1,z_2,d_1,d_2}(x_1,x_2) \Bigg[ w_{z_1,s}(x_1)w_{z_2,s}(x_2) + \frac{\{1 - w_{z_1,s}(x_1)\}w_{z_2,s}(x_2)}{\BF_{U[1]}^{z_1,z_2,d_1,d_2}(x_1,x_2)} \displaybreak[0]\\
        &\qquad\qquad\qquad\qquad\qquad\qquad + \frac{w_{z_1,s}(x_1)\{1-w_{z_2,s}(x_2)\}}{\BF_{U[2]}^{z_1,z_2,d_1,d_2}(x_1,x_2)} + \frac{\{1 - w_{z_1,s}(x_1)\}\{1-w_{z_2,s}(x_2)\}}{\BF_{U[1,2]}^{z_1,z_2,d_1,d_2}(x_1,x_2)} \Bigg]^{-1}, \displaybreak[0]\\
        &\qquad\qquad1-\{1-m_{z_1,z_2,d_1,d_2}(x_1,x_2)\} \displaybreak[0]\\
        &\qquad\qquad\qquad\times\Big[ w_{z_1,s}(x_1)w_{z_2,s}(x_2) + \{1 - w_{z_1,s}(x_1)\}w_{z_2,s}(x_2) \BF_{U[1]}^{z_1,z_2,d_1,d_2}(x_1,x_2) \displaybreak[0]\\
        &\qquad\qquad\qquad\qquad + w_{z_1,s}(x_1)\{1 - w_{z_2,s}(x_2)\} \BF_{U[2]}^{z_1,z_2,d_1,d_2}(x_1,x_2) \displaybreak[0]\\
        &\qquad\qquad\qquad\qquad + \{1 - w_{z_1,s}(x_1)\}\{1 - w_{z_2,s}(x_2)\} \BF_{U[1,2]}^{z_1,z_2,d_1,d_2}(x_1,x_2)\Big]^{-1} \Bigg).
    \end{align*}
\end{theorem}

Bounds in Theorem \ref{thm:PGCE-general-bounds} follow a conceptually similar structure to those in Theorem \ref{thm:general-bounds} on the PCEs, as each bound given in Theorem \ref{thm:PGCE-general-bounds} also contains two components: one from directly applying \eqref{eq:PGCE-BF-1}--\eqref{eq:PGCE-BF-1-2}, and the other from applying \eqref{eq:PGCE-BF-1}--\eqref{eq:PGCE-BF-1-2} to $1-h\{Y_1(z_1),Y_2(z_2)\}$ so that the obtained final bounds stay valid, since $h\{Y_1(z_1),Y_2(z_2)\}\in[0,1]$. To obtain the covariate-adjusted bounds on the PGCE $\mu_{z_1,z_2,s}$, we can now integrate the conditional bounds in Theorem \ref{thm:PGCE-general-bounds} over the stratum-specific joint covariate distribution $F_{X_1 | S_1=s}(x_1)F_{X_2 | S_2=s}(x_2)$. 

By Theorem \ref{thm:PGCE-general-bounds}, we can also obtain $[L_{z_1,z_2,s}^\mathrm{GM}(x_1,x_2), U_{z_1,z_2,s}^\mathrm{GM}(x_1,x_2)]$, the version of the worst-case bounds \citep{Grilli2008} for $\mu_{z_1,z_2,s}(x_1,x_2)$ by letting all bounding factors $\BF_{U[1]}^{z_1,z_2,d_1,d_2}(x_1,x_2)$, $\BF_{U[2]}^{z_1,z_2,d_1,d_2}(x_1,x_2)$, and $\BF_{U[1,2]}^{z_1,z_2,d_1,d_2}(x_1,x_2) \to \infty$, where
\begin{align*} 
    L_{z_1,z_2,s}^\mathrm{GM}(x_1,x_2) &= \max\left\{ 0, 1 - \frac{1 - m_{z_1,z_2,d_1,d_2}(x_1,x_2)}{w_{z_1,s}(x_1)w_{z_2,s}(x_2)} \right\},\\ 
    U_{z_1,z_2,s}^\mathrm{GM}(x_1,x_2) &= \min\left\{ 1, \frac{m_{z_1,z_2,d_1,d_2}(x_1,x_2)}{w_{z_1,s}(x_1)w_{z_2,s}(x_2)} \right\}.
\end{align*}

Also, by \eqref{eq:PGCE-BF-relationship}, we can obtain the empirically convenient but conservative approximate of the bounds in Theorem \ref{thm:PGCE-general-bounds} by replacing $\BF_{U[1]}^{z_1,z_2,d_1,d_2}(x_1,x_2)$ and $\BF_{U[2]}^{z_1,z_2,d_1,d_2}(x_1,x_2)$ with $\BF_{U[1,2]}^{z_1,z_2,d_1,d_2}(x_1,x_2)$, i.e.,
\begin{align*}
    &L_{z_1,z_2,s}^\dagger(x_1,x_2) \displaybreak[0]\\
    &= \max\Bigg( m_{z_1,z_2,d_1,d_2}(x_1,x_2) \Big[ w_{z_1,s}(x_1)w_{z_2,s}(x_2) + \{1 - w_{z_1,s}(x_1)w_{z_2,s}(x_2)\} \BF_{U[1,2]}^{z_1,z_2,d_1,d_2}(x_1,x_2)\Big]^{-1}, \displaybreak[0]\\
    &\qquad\qquad 1- \{1-m_{z_1,z_2,d_1,d_2}(x_1,x_2)\}\Bigg[ w_{z_1,s}(x_1)w_{z_2,s}(x_2) + \frac{1 - w_{z_1,s}(x_1)w_{z_2,s}(x_2)}{\BF_{U[1,2]}^{z_1,z_2,d_1,d_2}(x_1,x_2)} \Bigg]^{-1} \Bigg)
\end{align*}
and
\begin{align*}
    &U_{z_1,z_2,s}^\dagger(x_1,x_2) \displaybreak[0]\\
    &= \min\Bigg( m_{z_1,z_2,d_1,d_2}(x_1,x_2) \Bigg[ w_{z_1,s}(x_1)w_{z_2,s}(x_2) + \frac{1 - w_{z_1,s}(x_1)w_{z_2,s}(x_2)}{\BF_{U[1,2]}^{z_1,z_2,d_1,d_2}(x_1,x_2)} \Bigg]^{-1}, \displaybreak[0]\\
    &\qquad\qquad1-\{1-m_{z_1,z_2,d_1,d_2}(x_1,x_2)\} \displaybreak[0]\\
    &\qquad\qquad\qquad\times\Big[ w_{z_1,s}(x_1)w_{z_2,s}(x_2) + \{1 - w_{z_1,s}(x_1)w_{z_2,s}(x_2)\} \BF_{U[1,2]}^{z_1,z_2,d_1,d_2}(x_1,x_2)\Big]^{-1} \Bigg).
\end{align*}
The bounds $[L_{z_1,z_2,s}^\dagger(x_1,x_2),U_{z_1,z_2,s}^\dagger(x_1,x_2)]$ are generally not sharp unless the MPI assumption is not violated or the confounding is in the worst-case scenario. (Detailed explanations are given in Section S2.3 of the Supplementary Materials.) Corollary \ref{coro:PGCE-nested-bounds} summarizes the relationships among these three sets of bounds.

\begin{corollary} \label{coro:PGCE-nested-bounds}
    Under Assumptions \ref{asp:treatment-ignorability}--\ref{asp:strata-identification}, for any $ \BF_{U[1]}^{z_1,z_2,d_1,d_2}(x_1,x_2)$, $\BF_{U[2]}^{z_1,z_2,d_1,d_2}(x_1,x_2)$, and $\BF_{U[1,2]}^{z_1,z_2,d_1,d_2}(x_1,x_2) \in (1,\infty)$, we have $L_{z_1,z_2,s}^\mathrm{GM}(x_1,x_2) < L_{z_1,z_2,s}^\dagger(x_1,x_2) < L_{z_1,z_2,s}(x_1,x_2) \leq U_{z_1,z_2,s}(x_1,x_2) < U_{z_1,z_2,s}^\dagger(x_1,x_2) < U_{z_1,z_2,s}^\mathrm{GM}(x_1,x_2)$.
    As $ \BF_{U[1]}^{z_1,z_2,d_1,d_2}(x_1,x_2)$, $\BF_{U[2]}^{z_1,z_2,d_1,d_2}(x_1,x_2)$, and $\BF_{U[1,2]}^{z_1,z_2,d_1,d_2}(x_1,x_2) \to \infty$, $L_{z_1,z_2,s}(x_1,x_2) \downarrow L_{z_1,z_2,s}^\dagger(x_1,x_2) \downarrow L_{z_1,z_2,s}^\mathrm{GM}(x_1,x_2)$ and $U_{z_1,z_2,s}(x_1,x_2) \uparrow U_{z_1,z_2,s}^\dagger(x_1,x_2) \uparrow U_{z_1,z_2,s}^\mathrm{GM}(x_1,x_2)$.
\end{corollary}

Corollary \ref{coro:PGCE-nested-bounds} states that, if all bounding factors are finite, $[L_{z_1,z_2,s}(x_1,x_2), U_{z_1,z_2,s}(x_1,x_2)]$ nests within $[L_{z_1,z_2,s}^\dagger(x_1,x_2), U_{z_1,z_2,s}^\dagger(x_1,x_2)]$, which, in turn, nests within $[L_{z_1,z_2,s}^\mathrm{GM}(x_1,x_2),\allowbreak U_{z_1,z_2,s}^\mathrm{GM}(x_1,x_2)]$; if all bounding factors approach infinity in the worst-case scenario, then both $[L_{z_1,z_2,s}(x_1,x_2), U_{z_1,z_2,s}(x_1,x_2)]$ and $[L_{z_1,z_2,s}^\dagger(x_1,x_2), U_{z_1,z_2,s}^\dagger(x_1,x_2)]$ converge to $[L_{z_1,z_2,s}^\mathrm{GM}(x_1,x_2), U_{z_1,z_2,s}^\mathrm{GM}(x_1,x_2)]$. 

\subsection{Bounds under monotonicity}

In the special case of monotonicity, we have $e_{01}(x)=0$, which leads to $\calS_{1,1} = \{10, 11\}$, $\calS_{0,0} = \{10, 00\}$, $\calS_{1,0} = \{00\}$, and $\calS_{0,1} = \{11\}$. If both $\calS_{z_1,d_1}$ and $\calS_{z_2,d_2}$ are point identification sets, i.e., $(z_1,d_1),(z_2,d_2)\in\{(1,0),(0,1)\}$, then $\mu_{z_1,z_2,s}(x_1,x_2)$ is point identifiable with $s\in\calS_{z_1,d_1}\bigcap\calS_{z_2,d_2}$. If neither $\calS_{z_1,d_1}$ nor $\calS_{z_2,d_2}$ is a point identification set, i.e., $(z_1,d_1),(z_2,d_2)\in\{(1,1),(0,0)\}$, then the bounds on $\mu_{z_1,z_2,s}(x_1,x_2)$ are in the same form as the general bounds given in Theorem \ref{thm:PGCE-general-bounds}. If only one of $\calS_{z_1,d_1}$ and $\calS_{z_2,d_2}$ is a point identification set---without loss of generality, suppose that $\calS_{z_1,d_1}$ is a point identification set---then the bounds for $\mu_{z_1,z_2,s}(x_1,x_2)$ are given in Corollary \ref{coro:PGCE-monotone-bounds}.

\begin{corollary} \label{coro:PGCE-monotone-bounds}
    Under Assumptions \ref{asp:treatment-ignorability}--\ref{asp:strata-identification} with $\theta(x)\to\infty$, for $s\in\calS_{z_1,d_1}\bigcap\calS_{z_2,d_2}$, $(z_1,d_1)\in\{(1,0),(0,1)\}$, and $(z_2,d_2)\in\{(1,1),(0,0)\}$, $\mu_{z_1,z_2,s}(x_1,x_2)$ is sharply bounded by $L_{z_1,z_2,s}(x_1,x_2)\leq \mu_{z_1,z_2,s}(x_1,x_2) \leq U_{z_1,z_2,s}(x_1,x_2)$, where
    \begin{align*}
        &L_{z_1,z_2,s}(x_1,x_2) \displaybreak[0]\\
        &= \max\Bigg( m_{z_1,z_2,d_1,d_2}(x_1,x_2) \Big[ w_{z_2,s}(x_2) + \{1 - w_{z_2,s}(x_2)\} \BF_{U[2]}^{z_1,z_2,d_1,d_2}(x_1,x_2) \Big]^{-1}, \displaybreak[0]\\
        &\qquad\qquad 1- \{1-m_{z_1,z_2,d_1,d_2}(x_1,x_2)\} \Bigg\{ w_{z_2,s}(x_2) + \frac{1-w_{z_2,s}(x_2)}{\BF_{U[2]}^{z_1,z_2,d_1,d_2}(x_1,x_2)} \Bigg\}^{-1} \Bigg)
    \end{align*}
    and
    \begin{align*}
        &U_{z_1,z_2,s}(x_1,x_2) \displaybreak[0]\\
        &= \min\Bigg( m_{z_1,z_2,d_1,d_2}(x_1,x_2) \Bigg\{ w_{z_2,s}(x_2) + \frac{1-w_{z_2,s}(x_2)}{\BF_{U[2]}^{z_1,z_2,d_1,d_2}(x_1,x_2)}\Bigg\}^{-1}, \displaybreak[0]\\
        &\qquad\qquad1-\{1-m_{z_1,z_2,d_1,d_2}(x_1,x_2)\}\Big[ w_{z_2,s}(x_2) + \{1 - w_{z_2,s}(x_2)\} \BF_{U[2]}^{z_1,z_2,d_1,d_2}(x_1,x_2)\Big]^{-1} \Bigg).
    \end{align*}
\end{corollary}

Corollary \ref{coro:PGCE-monotone-bounds} is an implication of Theorem \ref{thm:PGCE-general-bounds}, as, essentially, $w_{z_1,s}(x_1)=1$ if $\calS_{z_1,d_1}$ is a point identification set. Furthermore, under monotonicity, the bounds only depend on a single bounding factor and obviate the need to elicit $\BF_{U[1]}^{z_1,z_2,d_1,d_2}(x_1,x_2)$ and $\BF_{U[1,2]}^{z_1,z_2,d_1,d_2}(x_1,x_2)$. By symmetry, we can also derive bounds on $\mu_{z_1,z_2,s}(x_1,x_2)$ in the case where $\calS_{z_2,d_2}$ is a point identification set, whereas $\calS_{z_1,d_1}$ is not. Similar to the general case, we can easily obtain $[L_{z_1,z_2,s}^\mathrm{LH}(x_1,x_2), U_{z_1,z_2,s}^\mathrm{LH}(x_1,x_2)]$, the worst-case bounds \citep{Long2013} on $\mu_{z_1,z_2,s}(x_1,x_2)$ under monotonicity by letting all constituent bounding factors approach infinity, where $[L_{z_1,z_2,s}^\mathrm{LH}(x_1,x_2),L_{z_1,z_2,s}^\mathrm{LH}(x_1,x_2)]$ is in the same form as $[L_{z_1,z_2,s}^\mathrm{GM}(x_1,x_2),L_{z_1,z_2,s}^\mathrm{GM}(x_1,x_2)]$ if $(z_1,d_1),(z_2,d_2)\in\{(1,1),(0,0)\}$, and 
\begin{align*} 
    L_{z_1,z_2,s}^\mathrm{LH}(x_1,x_2) &= \max\left\{ 0, 1 - \frac{1 - m_{z_1,z_2,d_1,d_2}(x_1,x_2)}{w_{z_2,s}(x_2)} \right\},\\ 
    U_{z_1,z_2,s}^\mathrm{LH}(x_1,x_2) &= \min\left\{ 1, \frac{m_{z_1,z_2,d_1,d_2}(x_1,x_2)}{w_{z_2,s}(x_2)} \right\},
\end{align*}
if $(z_1,d_1)\in\{(1,0),(0,1)\}$ and $(z_2,d_2)\in\{(1,1),(0,0)\}$ and vice versa. The empirically convenient conservative approximate is only useful if neither $\calS_{z_1,d_1}$ nor $\calS_{z_2,d_2}$ is a point identification set, where all three bounding factors are present, and the form of the approximate bounds under monotonicity is the same as $[L_{z_1,z_2,s}^\dagger(x_1,x_2),U_{z_1,z_2,s}^\dagger(x_1,x_2)]$ given in the general case. Similarly, $[L_{z_1,z_2,s}^\dagger(x_1,x_2),U_{z_1,z_2,s}^\dagger(x_1,x_2)]$ are generally not sharp unless the MPI assumption is not violated or the confounding is in the worst-case scenario. Corollary \ref{coro:PGCE-monotone-nested-bounds} summarizes the relationships among these three sets of bounds.

\begin{corollary} \label{coro:PGCE-monotone-nested-bounds}
    Under Assumptions \ref{asp:treatment-ignorability}--\ref{asp:strata-identification} and $\theta(x)\to\infty$, for any $ \BF_{U[1]}^{z_1,z_2,d_1,d_2}(x_1,x_2)$, $\BF_{U[2]}^{z_1,z_2,d_1,d_2}(x_1,x_2)$, and $\BF_{U[1,2]}^{z_1,z_2,d_1,d_2}(x_1,x_2) \in (1,\infty)$, we have $L_{z_1,z_2,s}^\mathrm{LH}(x_1,x_2) < L_{z_1,z_2,s}^\dagger(x_1,x_2) < L_{z_1,z_2,s}(x_1,x_2) \leq U_{z_1,z_2,s}(x_1,x_2) < U_{z_1,z_2,s}^\dagger(x_1,x_2) < U_{z_1,z_2,s}^\mathrm{LH}(x_1,x_2)$.
    As $ \BF_{U[1]}^{z_1,z_2,d_1,d_2}(x_1,x_2)$, $\BF_{U[2]}^{z_1,z_2,d_1,d_2}(x_1,x_2)$, and $\BF_{U[1,2]}^{z_1,z_2,d_1,d_2}(x_1,x_2) \to \infty$, $L_{z_1,z_2,s}(x_1,x_2) \downarrow L_{z_1,z_2,s}^\dagger(x_1,x_2) \downarrow L_{z_1,z_2,s}^\mathrm{LH}(x_1,x_2)$ and $U_{z_1,z_2,s}(x_1,x_2) \uparrow U_{z_1,z_2,s}^\dagger(x_1,x_2) \uparrow U_{z_1,z_2,s}^\mathrm{LH}(x_1,x_2)$.
\end{corollary}

\subsection{Cornfield connection}

Finally, we establish the Cornfield connection in the PGCE setting. Let $c_0 \in [0,1)$ be a pre-specified scientific null threshold for $\mu_{z_1,z_2,s}(x_1,x_2)$. For example, if $h(u,v) = \bbone(u > v)$, the null threshold indicating stochastic equivalence is typically $c_0 = 1/2$. Let $\rho_{SU}(x_1,x_2) = \max\{ \RR_{SU}^{z_1,d_1}(x_1), \RR_{SU}^{z_2,d_2}(x_2) \}$ and $\rho_{UY}(x_1,x_2) = \RR_{UY}^{z_1,z_2}(x_1,x_2)$. Define $L_{z_1,z_2,s}(x_1,x_2,\rho_{SU},\rho_{UY})$ and $U_{z_1,z_2,s}(x_1,x_2,\rho_{SU},\rho_{UY})$ respectively denote the bounds $L_{z_1,z_2,s}(x_1,x_2)$ and $U_{z_1,z_2,s}(x_1,x_2)$ with $\RR_{SU}^{z_1,d_1}(x_1)$ and $\RR_{SU}^{z_2,d_2}(x_2)$ replaced by $\rho_{SU}(x_1,x_2)$ and $\RR_{UY}^{z_1,z_2}(x_1,x_2)$ replaced by $\rho_{UY}(x_1,x_2)$. Similar to the PCE setting, without loss of generality, assume an apparently causal effect, where the point estimate of $\mu_{z_1,z_2,s}(x_1,x_2)$ under the MPI assumption exceeds $c_0$. Provided the worst-case lower bound crosses the null ($\lim_{\rho_{SU}(x_1,x_2), \rho_{UY}(x_1,x_2) \to \infty} L_{z_1,z_2,s}(x_1,x_2,\rho_{SU}, \rho_{UY}) \leq c_0$), the critical falsification thresholds are point identified as the unique real roots of three equations on the interval $(1, \infty)$. 

Specifically, the global maximum threshold $\Gamma_s^*(x_1,x_2)$ is defined as the root of the equation 
\begin{align*}
    L_{z_1,z_2,s}(x_1,x_2,\Gamma_s^*, \Gamma_s^*) = c_0.
\end{align*}
The outcome bias lower threshold $\gamma_{UY,s}^*(x_1,x_2)$ is identified as the root of the asymptotic limit where selection bias diverges ($\rho_{SU}(x_1,x_2) \to \infty$); mathematically, this is equivalent to solving $L_{z_1,z_2,s}(x_1,x_2,\gamma_{UY,s}^*, \gamma_{UY,s}^*) = c_0$ for $\gamma_{UY,s}^*(x_1,x_2)$. Conversely, the selection bias lower threshold $\gamma_{SU,s}^*(x_1,x_2)$ is identified as the root of the asymptotic limit where outcome bias diverges ($\rho_{UY}(x_1,x_2) \to \infty$); because the independent replicates compound multiplicatively, this is equivalent to solving $L_{z_1,z_2,s}(x_1,x_2,\gamma_{SU,s}^*, \gamma_{SU,s}^{*2}) = c_0$ for $\gamma_{SU,s}^*(x_1,x_2)$. Theorem \ref{thm:PGCE-cornfield} below gives the Cornfield-type conditions in the PGCE setting.

\begin{theorem} \label{thm:PGCE-cornfield}
    Under Assumptions \ref{asp:treatment-ignorability}--\ref{asp:strata-identification}, suppose the point estimate of $\mu_{z_1,z_2,s}(x_1,x_2)$ under the MPI assumption exceeds $c_0$ and the worst-case lower bound $L_{z_1,z_2,s}^\mathrm{GM}(x_1,x_2) \leq c_0$. There exist three unique critical falsification thresholds, $1 < \gamma_{SU,s}^*(x_1,x_2) < \gamma_{UY,s}^*(x_1,x_2) < \Gamma_s^*(x_1,x_2) < \infty$, such that to nullify $\mu_{z_1,z_2,s}(x_1,x_2)$ (reduce $L_{z_1,z_2,s}(x_1,x_2)$ to $c_0$ or below), the latent unmeasured confounding must simultaneously satisfy the following Cornfield-type conditions:
    \begin{align}
        &\max\left\{ \RR_{SU}^{z_1,d_1}(x_1), \RR_{SU}^{z_2,d_2}(x_2) \right\} \geq \gamma_{SU,s}^*(x_1,x_2), \quad \RR_{UY}^{z_1,z_2}(x_1,x_2) \geq \gamma_{UY,s}^*(x_1,x_2), \label{eq:PGCE-cornfield-condition-1}\displaybreak[0]\\
        &\max\left\{ \RR_{SU}^{z_1,d_1}(x_1), \RR_{SU}^{z_2,d_2}(x_2), \RR_{UY}^{z_1,z_2}(x_1,x_2)\right\} \geq \Gamma_s^*(x_1,x_2). \label{eq:PGCE-cornfield-condition-2}
    \end{align}
\end{theorem}

Theorem \ref{thm:PGCE-cornfield} establishes the joint magnitude of unmeasured confounding required to nullify the target PGCE. The Cornfield-type conditions in Theorem \ref{thm:PGCE-cornfield} are analogous in structure to those in Theorem \ref{thm:cornfield}. Specifically, in the PCE setting, the thresholds satisfy $\gamma_{UY,s}^*(x_1,x_2)=\gamma_{UY}^*(x_1,x_2)$, and hence \eqref{eq:PGCE-cornfield-condition-1} reduces to \eqref{eq:cornfield-condition-1} with one selection RR. Consequently, \eqref{eq:PGCE-cornfield-condition-2} also reduces to \eqref{eq:cornfield-condition-2} in the PCE setting. 

When the target estimand is the PCGE, the Cornfield-type conditions are generally more involved. In particular, the thresholds $\gamma_{SU,s}^*(x_1,x_2)$ and $\gamma_{UY,s}^*(x_1,x_2)$ represent the mandatory minima for the unmeasured confounding. If one can theoretically bound the maximum unmeasured selection RR across both replicates to be less than $\gamma_{SU,s}^*(x_1,x_2)$, then $\mu_{z_1,z_2,s}(x_1,x_2)$ is immune against falsification, even if the unmeasured confounder perfectly dictates the joint outcome ($\RR_{UY}^{z_1,z_2}(x_1,x_2) \to \infty$). An identical, independent guarantee holds if the outcome RR is bounded below $\gamma_{UY,s}^*(x_1,x_2)$. The inequality $\gamma_{SU,s}^*(x_1,x_2) < \gamma_{UY,s}^*(x_1,x_2)$ formalizes that $\mu_{z_1,z_2,s}(x_1,x_2)$ is structurally more vulnerable to selection confounding than to outcome confounding. This is because the replicates operate independently, the unmeasured selection penalties compound multiplicatively within the joint nuisance cross-stratum $(s_1',s_2')$, whereas the shared outcome penalty remains linear. The universal maximum threshold $\Gamma_s^*(x_1,x_2)$ provides the one-number Cornfield-type summary metric. It shows that regardless of the specific algebraic balance between the selection and outcome bias magnitudes, the strongest latent confounding pathway must exceed $\Gamma_s^*(x_1,x_2)$ to force the lower bound to $c_0$. This serves as the conservative, univariate reporting metric for the robustness of the PGCE.

Similar to the principal E-value in \eqref{eq:e-value}, the maximum threshold derived in \eqref{eq:PGCE-cornfield-condition-2} formally motivates the \emph{generalized principal E-value} in the PGCE setting. That is, given an estimated effect strictly greater than $c_0$, the \emph{generalized principal E-value} is defined as the symmetric contour vertex, such that
\begin{align} \label{eq:PGCE-e-value}
    \mathrm{EV}_s(x_1,x_2) = \Gamma_s^*(x_1,x_2).
\end{align}
This metric quantifies the absolute minimum magnitude required for the maximum of the independent selection RRs and the outcome RR to fully nullify the estimated PGCE. 

Finally, Proposition \ref{prop:E-value-monotone-theta} further details the relationship between the odds ratio sensitivity parameter $\theta(x)$ and the E-values in \eqref{eq:e-value} and \eqref{eq:PGCE-e-value} for the $10$ stratum (e.g, compliers in the noncompliance scenario), and establishes the worst-case boundary for falsification.

\begin{proposition} \label{prop:E-value-monotone-theta}
    Suppose $p_1(x) > p_0(x)$. Let $\Gamma_{10}^*(\bftheta,\bfx)$ denote the principal E-value for stratum $10$ evaluated over $\Theta$ or $\Theta\times\Theta$. For \eqref{eq:e-value} in the PCE setting, $\bftheta = \theta(x) \in \Theta$ and $\bfx = x$. For \eqref{eq:PGCE-e-value} in the PGCE setting, $\bftheta = (\theta(x_1), \theta(x_2)) \in \Theta\times\Theta$ and $\bfx = (x_1,x_2)$. Assume the nonparametric baseline estimate under the MPI assumption exceeds $c_0$. Then, $\Gamma_{10}^*(\bftheta,\bfx)$ is strictly monotonically decreasing with respect to the product order on $\Theta$. The global infimum is uniquely achieved at the limit $\bftheta \to \bm\infty$ (i.e., monotonicity), establishing the worst-case structural boundary for falsification.
\end{proposition}

Interestingly, Proposition \ref{prop:E-value-monotone-theta} shows that the monotonicity assumption creates a worst-case adversarial setting for the falsification against the PCEs and PGCEs in the $10$ stratum (same for the $01$ stratum under reverse monotonicity). This is because, under monotonicity ($p_1(x) > p_0(x)$), evaluating the limit $\bftheta \to \bm\infty$ forces $e_{01}(x) \to 0$, which minimizes the $10$ proportion $e_{10}(x)$, since $e_{10}(x) = p_1(x) - p_0(x) + e_{01}(x)$. However, by the marginal constraints $e_{11}(x) = p_0(x) - e_{01}(x)$ and $e_{00}(x) = 1 - p_1(x) - e_{01}(x)$, minimizing $e_{01}(x)$ maximizes the proportions of the $11$ and $00$ strata. Therefore, the partial derivatives of their mixing weights $w_{z,11}$ and $w_{z,00}$ with respect to $\bftheta$ are positive. Subsequently, the E-values for the $11$ and $00$ strata are strictly monotonically increasing with respect to $\bftheta$. For these strata, the monotonicity assumption leads to the best-case (e.g., most robust) structural boundary for falsification.

\section{Data examples} \label{sec:illustration}

\subsection{Example one: the WHO-LARES study}

Our first data example is the World Health Organization's Large Analysis and Review of European Housing and Health Status (WHO-LARES) study with 5,882 individuals for the effect of living in damp or moldy conditions ($Z=1$ if yes and 0 if no) on depression ($Y=1$ if yes and 0 if no), where dampness or mold-related diseases is the intermediate outcome ($D=1$ if yes and 0 if no) \citep{VanderWeele2010}. For the purpose of illustration, we let $X=x\in\{0,1\}$ denote whether the subject has received secondary education or above (low vs. high education). Our target estimand is the principal causal relative risk $\mathrm{PCRR}_{10}(x)$ among the harmed population. That is, the subpopulation that would be diseased only if living in damp or moldy conditions. 

\begin{figure}[ht]
    \centering
    \includegraphics{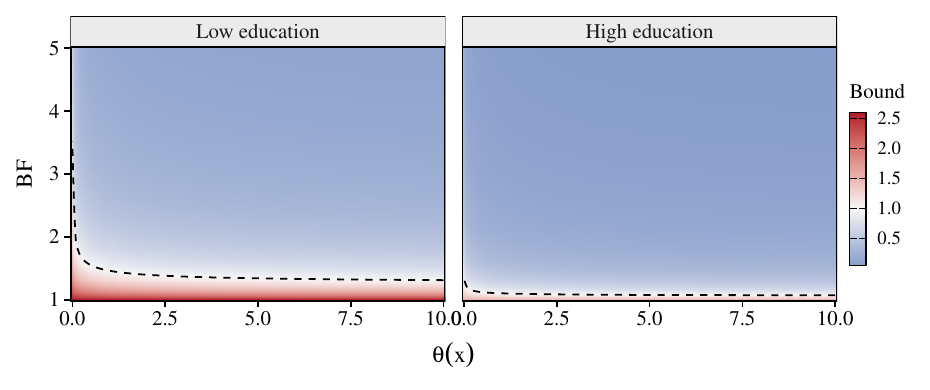}
    \caption{Illustration of the sensitivity lower bound surfaces for $\mathrm{PCRR}_{10}(x)$. The dashed lines depict the $\BF(x)$ that satisfies $L_{1,10}(x,\BF)/U_{0,10}(x,\BF) = 1$ for each $\theta(x)$. That is, the bounding factor required to drive the lower bound towards the null.}
    \label{fig:PCE-surface}
\end{figure}

Under the MPI assumption (i.e., $\BF=1$), for stratum $10$, the empirical nonparametric estimates of the causal relative risk are $2.60$ and $1.55$ for the low and high education groups, respectively. Because these baseline estimates exceed $1$, we deploy the sensitivity lower bound to quantify the robustness of this finding. Figure \ref{fig:PCE-surface} presents the sensitivity lower bounds for $\mathrm{PCRR}_{10}(x)$, each of which is a surface with two sensitivity parameters $\BF$ and $\theta$. Figures illustrating the sensitivity bound surfaces for the remaining three strata are provided in Section S3.1 of the Supplementary Materials.

\begin{figure}[ht]
  \centering
  \begin{subfigure}[t]{0.48\textwidth}
    \centering
    \includegraphics[width=\linewidth]{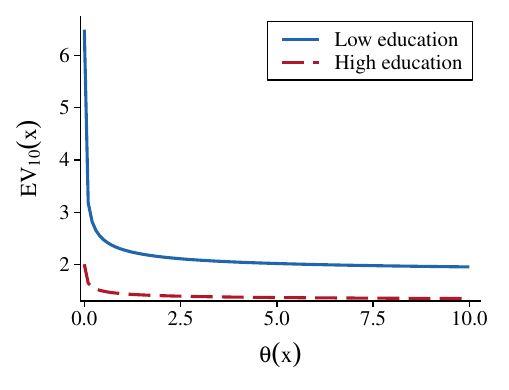}
    \caption{Principal E-value across $\theta(x)$.}
    \label{fig:PCE-evalue}
  \end{subfigure}
  \hfill
  \begin{subfigure}[t]{0.48\textwidth}
    \centering
    \includegraphics[width=\linewidth]{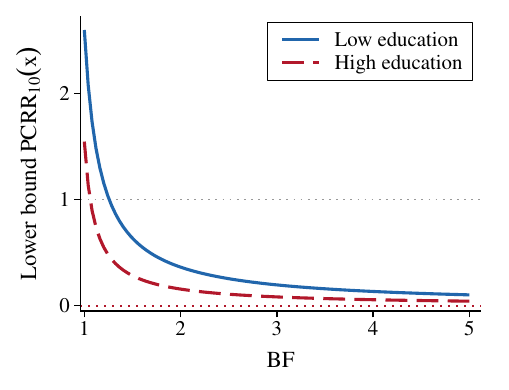}
    \caption{PCRR lower bound across $\BF$.}
    \label{fig:PCE-bounds}
  \end{subfigure}

  \vspace{0.8em}

  \begin{subfigure}[t]{0.48\textwidth}
    \centering
    \includegraphics[width=\linewidth]{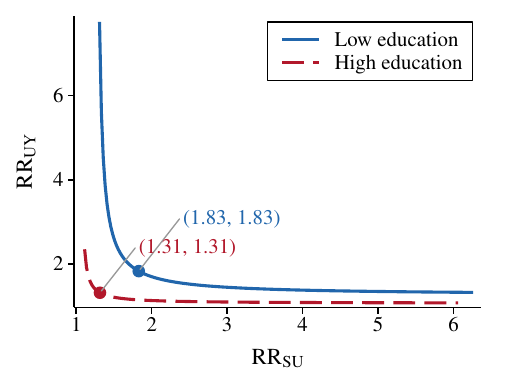}
    \caption{Cornfield falsification contour under monotonicity.}
    \label{fig:PCE-contour}
  \end{subfigure}
  \hfill
  \begin{subfigure}[t]{0.48\textwidth}
    \centering
    \includegraphics[width=\linewidth]{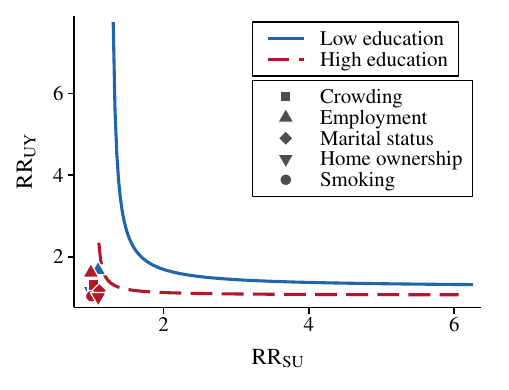}
    \caption{Falsification map with empirical benchmarks.}
    \label{fig:PCE-benchmarks}
  \end{subfigure}
  \caption{Illustration via the WHO-LARES example. 
    (a) Principal E-value curve; 
    (b) PCRR lower bound; 
    (c) joint confounding contour; 
    (d) contour with measured-covariate benchmarks.}
  \label{fig:PCE-sensitivity-panel}
\end{figure}

Figure \ref{fig:PCE-evalue} plots the E-value curves for the two groups with different values of $\theta(x)$, where as $\theta(x) \to \infty$, the E-value of the complier PCRR decreases, confirming Proposition \ref{prop:E-value-monotone-theta}. The E-value curves of the remaining three strata are given in Section S3.1 of the Supplementary Materials. Since monotonicity gives the worst-case structural boundary for falsification, we focus on this limit; any causal claim that survives the $\theta(x) \to \infty$ boundary holds for any finite $\theta(x)$. Figure \ref{fig:PCE-bounds} presents the sensitivity lower bounds for $\mathrm{PCRR}_{10}(x)$ under monotonicity. As the bounding factor increases, the lower bounds monotonically decrease and eventually cross $1$ from above, demonstrating that the observed causal effect is not immune to unmeasured confounding between $S$ and $Y(z)$ and can be nullified. The worst-case lower bounds for both groups are $0$, which is given as the darker dotted line, confirming that the proposed lower bound is above the worst-case one.

Figure \ref{fig:PCE-contour} shows the Cornfield contours under monotonicity, along with the high thresholds for both RRs. To demonstrate the utility, we computed empirical benchmarks of $\RR_{SU}(x)$ and $\RR_{UY}(x)$ using observed covariates in place of the unobserved confounder, given in Figure \ref{fig:PCE-benchmarks}. The observed covariates include employment, smoking, marital status, ownership, and crowding, and none of the empirical benchmark points lie above the Cornfield contours for either group. This indicates that any unmeasured confounding mimicking the measured confounding in the WHO-LARES data is insufficient to nullify $\mathrm{PCRR}_{10}(x)$.

\subsection{Example two: the U.S. Job Corps study}

Our second data example is the U.S. Job Corps study with 9,240 individuals for the effect of the training programs offered at Jobs Corps centers ($Z=1$ if yes and 0 if no) on earnings two years after the initial assignment ($Y$), where $Y\in\{0,1,2\}$ is ordinal, with $Y=0$ for unemployment, $Y=1$ for an hourly wage between \$0 and \$4.25, and $Y=2$ for above \$4.25, following the categorization considered in \citet{Chen2026}. The intermediate variable is the compliance with the assignment status one year after the initial assignment ($D=1$ if yes and 0 if no) \citep{Schochet2001}. We let $X=x\in\{0,1\}$ denote whether the subject is female (yes vs. no). Our target estimand is the complier PGCE $\mu_{1,0,10}(x_1,x_2)$, where the contrast function $h(u,v) = \bbone(u>v)+\frac{1}{2}\bbone(u=v)$, and we let $x_1=x_2=x$ for the purpose of interpretability (within group comparisons only). 

\begin{figure}[ht]
    \centering
    \includegraphics{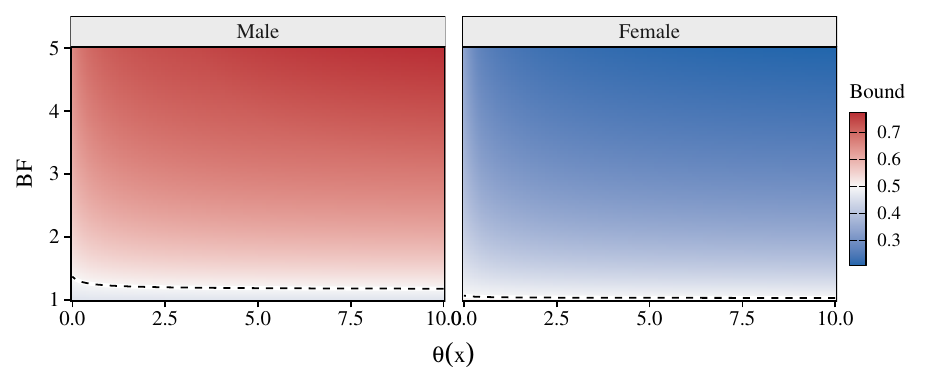}
    \caption{Illustration of the sensitivity bound surfaces for $\mu_{1,0,10}(x)$, where the upper bound is derived for the male group and the lower bound for the female group. The dashed line depicts the $\BF$ that satisfies $L_{1,0,10}(x,x,\sqrt{\BF},\BF) = 0.5$ for each $\theta$.}
    \label{fig:PGCE-surface}
\end{figure}

Under the MPI assumption, for stratum $10$ (compliers), the empirical nonparametric estimates of the PGCE are $0.46$ and $0.51$ for the male and female groups, respectively. We thus deploy the sensitivity upper bound to quantify the robustness of the finding for the male group, and the lower bound for the female group. Figure \ref{fig:PGCE-surface} presents the sensitivity bounds for $\mu_{1,0,10}(x)$. For simplicity and to aid visualization, the proposed bounds are plotted with $\BF_{U[1]} = \BF_{U[2]} = \sqrt{\BF_{U[1,2]}}$. Similar to Figure \ref{fig:PCE-surface}, each bound in Figure \ref{fig:PGCE-surface} is a surface with two sensitivity parameters $\BF$ and $\theta$ under this specification. Figures illustrating the sensitivity bound surfaces for the remaining three strata are provided in Section S3.2 of the Supplementary Materials.

\begin{figure}[ht]
  \centering
  \begin{subfigure}[t]{0.48\textwidth}
    \centering
    \includegraphics[width=\linewidth]{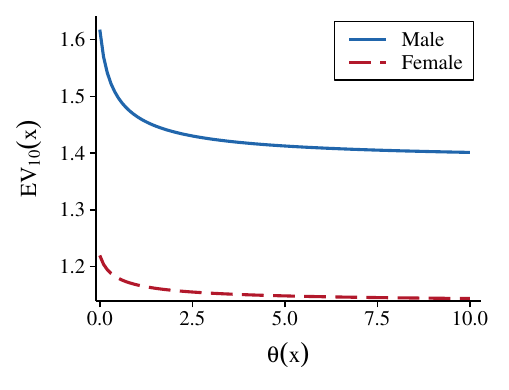}
    \caption{Generalized principal E-value across $\theta(x)$.}
    \label{fig:PGCE-evalue}
  \end{subfigure}
  \hfill
  \begin{subfigure}[t]{0.48\textwidth}
    \centering
    \includegraphics[width=\linewidth]{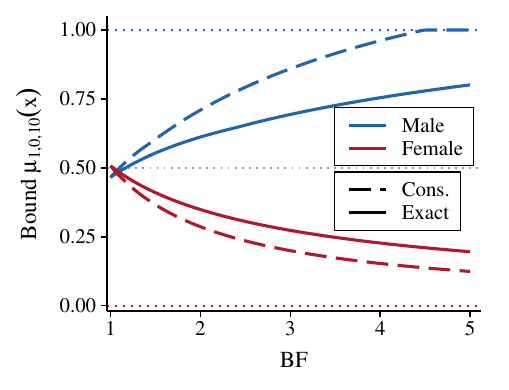}
    \caption{PGCE bounds across $\BF$.}
    \label{fig:PGCE-bounds}
  \end{subfigure}

  \vspace{0.8em}

  \begin{subfigure}[t]{0.48\textwidth}
    \centering
    \includegraphics[width=\linewidth]{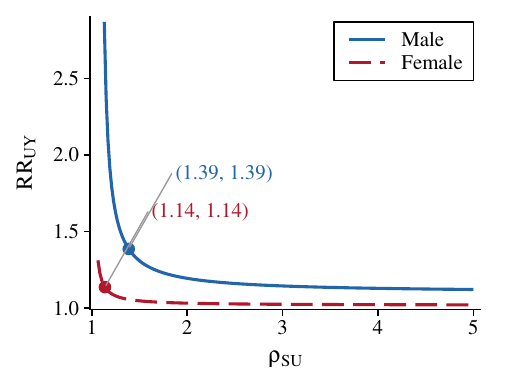}
    \caption{Cornfield falsification contour under monotonicity.}
    \label{fig:PGCE-contour}
  \end{subfigure}
  \hfill
  \begin{subfigure}[t]{0.48\textwidth}
    \centering
    \includegraphics[width=\linewidth]{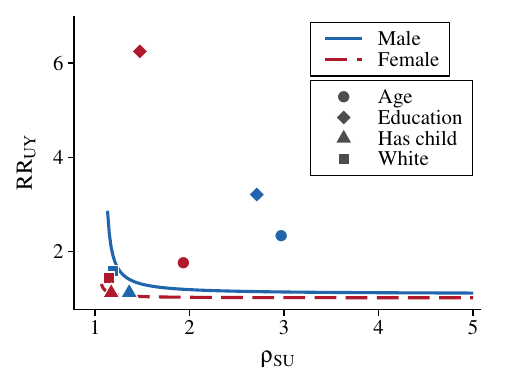}
    \caption{Falsification map with empirical benchmarks.}
    \label{fig:PGCE-benchmarks}
  \end{subfigure}
  \caption{Illustration via the U.S. Job Corps example. 
    (a) Generalized principal E-value curve; 
    (b) PCRR lower bound; 
    (c) joint confounding contour; 
    (d) contour with measured-covariate benchmarks.}
  \label{fig:PGCE-sensitivity-panel}
\end{figure}

Figure \ref{fig:PGCE-evalue} gives the generalized E-value curves for the two groups with different values of $\theta(x)$, where, similar to Figure \ref{fig:PCE-evalue}, as $\theta(x) \to \infty$, the E-value of the complier PGCE decreases. The generalized E-value curves of the remaining three strata are given in Section S3.2 of the Supplementary Materials. Figure \ref{fig:PGCE-bounds} presents the sensitivity bounds for $\mu_{1,0,10}(x)$ under monotonicity. As the bounding factor increases, the upper bounds monotonically increase and cross $0.5$ from below, and the lower bounds monotonically decrease and cross $0.5$ from above, demonstrating that the observed causal effect can be nullified by the unmeasured confounding between $S$ and $Y(z)$. The proposed bounds are nested within the conservative bounds, which in turn are nested within the worst-case bounds, represented by the darker dotted line.

Figure \ref{fig:PGCE-contour} shows the Cornfield contours under monotonicity, along with the high thresholds for the outcome RR and the maximum of the two selection RRs. We computed empirical benchmarks of $\RR_{UY}(x)$ and $\rho_{SU}(x)$ using observed covariates, given in Figure \ref{fig:PGCE-benchmarks}. The observed covariates include age, education, whether they have children, and whether they are white. For the male group, age and education give empirical benchmarks above the Cornfield contour. This indicates that the required magnitude of unmeasured confounding necessary to drive the upper bound to the null is less than the observed confounding strength of age and education. For the female group, all variables give empirical benchmarks above the Cornfield contour, with age and education showing the largest magnitudes. Because the estimated effect is highly sensitive to confounding magnitudes smaller than those of known, standard covariates, causal identification seems more structurally fragile under the MPI assumption. Any strong causal interpretation of the baseline estimates must therefore be heavily caveated, as a relatively moderate unmeasured confounder would nullify the observed PGCE estimate.

\section{Discussions} \label{sec:discussion} 

We developed a nonparametric sensitivity analysis framework for principal stratification that quantifies violations of the PI assumption via a margin-free bounding factor parameterized by the selection and outcome RRs of a set of unmeasured confounders. While the bounding factor has previously appeared in \citet{Ding2016epi} without an intermediate variable, the principal stratification setting introduces unique features that we address in this work. That is, the target stratum mean is latent and enters the observable cell mean as part of a weighted mixture whose mixing weights depend on the cross-world odds ratio parameter $\theta(x)$, creating a sensitivity dimension with no counterpart in the existing literature for nonparametric bounds. The sharp bounds in Theorems \ref{thm:general-bounds} and \ref{thm:PGCE-general-bounds} must simultaneously resolve this mixture decomposition and the bounding factor constraint. Interestingly, we proved that for any finite bounding factor, the proposed bounds nest strictly within the worst-case bounds of \citet{Grilli2008} and \citet{Long2013}, converging to them only as the bounding factor diverges. This nesting establishes a continuous algebraic bridge between point identification under PI and the boundary of maximum adversarial confounding, unifying disparate strands of the principal stratification literature. 

Beyond the traditional PCE, the PGCE estimand has been recognized as a useful target of causal inference for more complex endpoints \citep{Mao2018, mao2024wilcoxon}. The PGCE is particularly pertinent when the scientific objective focuses on the relative ranking of outcomes rather than on direct contrasts of mean responses—for example, in win–loss comparisons for ordinal endpoints in clinical trials, or in truncation-by-death settings, where the survivor probabilistic index offers a more robust and interpretable summary of treatment effect than the survivor average causal effect \citep{Chen2026}. The extension of our sensitivity framework to PGCEs with nonlinear contrast functions, therefore, addresses a pressing gap in the identification assumption. Compared to the PCE setting, this extension also introduces genuinely non-trivial structure: the non-separability of $h\{Y_1(z_1), Y_2(z_2)\}$ requires the outcome RR to be defined jointly over the unmeasured confounders of both replicates, producing three distinct bounding factors whose hierarchy reflects the multiplicative compounding of independent selection penalties in the product space, and the Cornfield-type analysis in Theorem \ref{thm:PGCE-cornfield} yields three distinct falsification thresholds whose ordering $\gamma_{SU,s}^*(x_1,x_2) < \gamma_{UY,s}^*(x_1,x_2) < \Gamma_s^*(x_1,x_2)$ formalizes that the pairwise estimand is structurally more vulnerable to selection confounding than to outcome confounding. This phenomenon is absent from both the PCE setting and prior sensitivity results for principal stratification.

We note that several directions merit future investigation. First, the current framework is established for a binary intermediate variable, which is the most typical setup for principal stratification. Extending the margin-free bounding parameters to continuous intermediate variables requires evaluating the confounding penalties over a continuous latent mixture, which involves substantially different identification arguments \citep{Lu2025, zhang2025semiparametric}. Second, the proposed bounds depend on the conditional observed cell means $m_{z,d}(x)$ and the principal scores $e_s(x)$, both of which are unknown and require estimation from the observed data. In high-dimensional covariate settings, model misspecification in these nuisance functions can distort the empirical bounds, and incorporating debiased machine learning estimators \citep{Levis2025, Tong2025} for estimation would provide formal guarantees on convergence rates and enable valid inference via confidence bands on the sensitivity bounds.

\section*{Acknowledgement}

This work is supported by the United States National Institutes of Health (NIH), National Heart, Lung, and Blood Institute (NHLBI, grant numbers R01-HL168202 and R01-HL178513). All statements in this report, including its findings and conclusions, are solely those of the authors and do not necessarily represent the views of the NIH.

\singlespacing
\bibliographystyle{jasa3}
\bibliography{Bounds}

\end{document}